\documentclass[aps,preprint,groupedaddress,reprint,showpacs,amsmath,amssymb,pre,onecolumn,12pt]{revtex4-1}
 \usepackage{graphicx}
 \usepackage{epstopdf, epsfig}
 \usepackage{xcolor}
 \usepackage[utf8]{inputenc}
 \usepackage[T1]{fontenc}
 \usepackage{lmodern}
 \usepackage{amsmath}
 \usepackage[lofdepth,lotdepth]{subfig}
 
\usepackage{soul}  

\begin{document}

\title{Miles' mechanism for generating surface water waves by wind, in finite water depth and subject to constant vorticity flow}

\author{N. Kern$^1$}
\author{C. Chaubet $^1$}
\author{R. A. Kraenkel$^2$}
\author{M. A. Manna$^1$}
\affiliation{$^1$Universit\'e Montpellier, Laboratoire Charles Coulomb UMR 5221, CNRS UM, France, F-34095, Montpellier, France.}
 \affiliation{$^2$Instituto de F\'isica T\'eorica-UNESP, Universidade Estadual Paulista, Rua Dr. Bento Teobaldo Ferraz 271 Bloco II, 01140-070, S\~ao Paulo, Brazil}

\pacs{}

\begin{abstract} 
The Miles' theory of wave amplification by wind is extended to the case of finite depth $h$ and a shear flow with (constant) vorticity $\Omega$. Vorticity is characterised through the non-dimensional parameter $\nu=\Omega U_1/g$, where $g$ the gravitational acceleration, $U_1 $ a characteristic wind velocity and $k$ the wavenumber.
The notion of 'wave age' is generalised to account for the effect of vorticity.
Several widely used growth rates are derived analytically from the dispersion relation of the wind/water interface, and their dependence on both water depth and vorticity is derived and discussed. 
Vorticity is seen to shift the maximum wave age, similar to what was previously known to be the effect of water depth. At the same time, a novel effect arises and the growth coefficients, at identical wave age and depth, are shown to experience a net increase or decrease according to the shear gradient in the water flow.
\end{abstract}

\maketitle

\section{Introduction : Wave generation by wind in water of finite depth} 
How wind generates ocean waves is a formidable problem, many aspects of which are still not understood today. It starts from the Navier-Stokes equations \cite{whitham}, applied to two layers of fluid. Solving these is a classic exercise, but the presence of wind requires a succession of subtle approximations and assumptions. In particular, the interface between water and air requires careful thought. The pioneering works are those due to Jeffreys  \cite{jeffreys1, jeffreys2} and to Miles \cite{miles}.
The theory by Jeffreys assumes that, on the lee-side of the surface wave, the ondulated water-air interface is sheltered from the air flowing over it surface waves. 
This kinematic scenario produces a pressure gradient which performs work on the wave. Hence energy is transferred from the wind to the wave (Jeffrey's {\it sheltering mechanism}).
Miles' theory of wave generation by wind assumes that ocean surface waves are generated by a resonance phenomenon. Resonance appears between the wave-induced pressure gradient on the inviscid airflow and the surface waves. This leads to a growing wave amplitude when the phase velocity of the surface wave equals the speed of the airflow (see reference \citep{Janssen} for a thorough discussion). In Jeffreys' and Miles' theories the viscosity is neglected both in air and in water, and furthermore water is considered as infinitely deep and irrotational. Both theories rely on linearising the equations of motions. 

Historically the first experiments and numerical studies concerning finite depth wind-wave growth were conducted by Thijsse  \cite{Thijsse} and Bretschneider \cite{Bretschneider}. Particularly important, in order to understand the physics of wind-wave dynamics in finite depth, were the experiments carried out in Lake George (Australia) by Young and Verhagen \cite{YoungVerhagen1,YoungVerhagen2}. For a full account of the Young and Verhagen works see reference \cite{Young3}.

Recently, in reference \cite{Manna1}, Miles' theory was extended to water of finite depth, and in references \cite{Manna2, Manna3} the wind-wave amplification was studied as a possible mechanism for finite-time blow-up of solitary wind waves and steep wave events (rogue waves) in finite water depth.
In reference \cite{Manna4} we have established a {\it fully nonlinear} model equation of surface wind-wave generation. There, the interaction between air and water is described from a quasi-linear point of view, in which the water obeys the fully nonlinear Serre-Green-Naghdi model \cite{GLN1, GLN2, GLN3} while, as in Miles' theory, the air flow is kept linear and obeys the linear Euler equation of motion. 

Net currents in the water are absent in all these approaches. In coastal waters, underlying currents may be present, and this raises the question how the propagation of wind-driven surface waves is affected.
Such currents may range from weak to very strong, and can be generated through various mechanisms such as tidal flow, oceanic circulation, wind action or breaking of waves. Depending on how they are generated the currents are often observed to vary with depth, and thus carry an underlying {\it vorticity}. 
Such background vorticity, known to be present for example in strong tidal currents \citep{Soulsby90} or in wind driven currents \citep{Jonsson1990}, can be important and should be taken into account when modeling 
the propagation of water waves \citep{TCRB16}.
Vorticity is especially observed in shallow water environments. 
For instance, linear shear due to strong currents has been observed in the surf zone, in strong rip currents, in situ \citep{MTSR05} or in laboratory experiments \citep{HS02}. More recently, a similar depth dependence of currents was observed over coral reefs \citep{SCDBP17}.

From a theoretical point of view, after the pioneering works of Benjamin  \citep{Benjamin} and Freeman and Johnson \citep{Freeman}, the role played by constant or variable vorticity (almost exponentially decreasing in depth) constitutes a classical but vast subject
in fluid mechanics. Many theories, exact or approximate, were formulated for steady periodic waves and for progressive waves, in finite or infinite depth, under the action of vorticity   \citep{Silva, Constan1, Constan2, Constan3, Constan4, Constan5}.
The combined action of vorticity and surface tension were studied in references \citep{Brantenberg} \citep{Kang}. The modulational instability (Benjamin-Feir instability) in finite depth under the action of vorticity was investigated in  \citep{Thomas}.

The aim of this work is to provide a theory for the growth of surface wind waves, in water of finite depth and in presence of constant vorticity currents. 
The purpose is twofold: on one hand it intends to establish the mathematical laws able to qualitatively reproduce at least some crucial features of the field experiments. On the other hand the intention is to supply a theoretical framework, thus going beyond empirical laws.

To carry out this task we build upon the theory by Miles in a finite depth context, as introduced in reference \citep{Manna1}, which we extend in order to account for the action of a linear shear current. 
The paper is organized as follows.

Section (\ref{sec:MathematicalAnalysis}) lays out the mathematical framework.
In subsection (\ref{water domain}) we introduce the nonlinear Euler equation governing the dynamics of surface water waves under the action of constant vorticity. 
In subsection (\ref{air domain}) the air domain is introduced and coupled to the water domain, leading to an air pressure which is no longer an overall constant, but instead varies in space and over time. Solving the linear problem at the interface we derive the linear dispersion relation of wave amplification, in finite depth and in presence of a constant shear in the flow field.
In section (\ref{wave growth}) we introduce dimensionless variables and scalings, to derive various adequate growth rates which are commonly considered in the field.
The resulting linear dynamics is described analytically, in terms of explicit results, which we then exploit numerically. Finally, section (\ref{sec:conclusions}) draws the conclusions to our findings.


\section{Mathematical analysis }
\label{sec:MathematicalAnalysis}

Here we lay out the equations describing the system consisting of a water domain (including the underlying shear current which induces vorticity) and the air domain (including the flow field due to wind), before coupling these at the interface on which a wave propagates.

\begin{figure}
    \centering
    \includegraphics[width=0.49\textwidth]{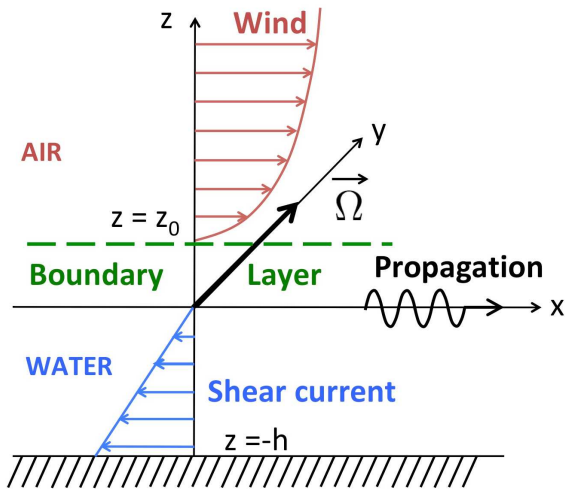}
   \caption{\label{fig:sketch}
    Sketch to illustrate the reference frame, the geometry as well as the flow conditions in water and in air. Both water currents and wind are taken towards the right. The example has positive vorticity, meaning that the current decreases with depth. The system is invariant in the $y$ direction.
    }
\end{figure}

In the absence of a propagating wave the situation is as follows, also represented in the sketch in Fig. (\ref{fig:sketch}). We consider the system to be invariant in the y-direction. Thus water and air particles are located in a two-dimensional Cartesian coordinate system with axes $x,z$. The origin is at $x=z=0$, and the horizontal domain is $x \in ]-\infty,+\infty[$. At a height $z=0$ is the water-air interface, the domain of positive $z\in]0,\infty[$ corresponds to the (unperturbed) air, and $z \in [-h,0]$ is the (unperturbed) water domain. The bottom of the water domain, at $z=-h$, is considered unpermeable. Both water and air are taken to be inviscid and incompressible, and surface tension effects at the interface are not accounted for. 
\\

We assume the water to flow in the positive $x$ direction, the velocity varying linearly with depth. In an earth-bound reference frame the velocities therefore interpolate between $v_s \,\vec e_x$ at the water surface ($z=0$) and $v_b \, \vec e_x$ at the bottom ($z=-h$),  where $\vec{e}_x$ is the unit vector. The surface velocity $v_s$, bottom velocity $v_b$ as well as the water depth $h$ are taken to be constants. In the following we will work in the reference frame in which the water surface is  static, i.e. the  frame which moves at the surface velocity $v_s \, \vec e_x$ with respect to the earth-bound reference frame.  This is a natural choice, since the interface is the support for the propagating wave, although there are some implications which we will return to later.

In the reference frame of the water surface the water flow profile due to vorticity is thus given by 
\begin{equation}\label{profile}
 \vec{U}^{(\Omega)}_{0}(z)= \Omega z \, \vec{e}_x \quad \mbox{for}\quad -h \leq z\leq 0
  \ ,
 \end{equation}
where the parameter
 \begin{equation}
 \Omega= \frac{v_s-v_b}{h}
 \end{equation}
 is in fact the (y-component of the)  vorticity attributed to the steady shear flow ($\vec{\omega} = \nabla \times \vec U^{(\Omega)}_0(z) = \Omega\, \vec{e}_y = const$).
\\

For the air domain, a steady-state airflow is prescribed to represent the effect of wind. To this end the mean horizontal velocity profile is of the form
\begin{equation}\label{airprofile}
\vec{U}(z) = U(z) \, \vec{e}_x = U_1 \, \hat{U}(z) \, \vec{e}_x
\ ,
\end{equation}
where $U_1$ is a characteristic wind speed and $\hat{U}$ is a dimensionless function characterising the wind profile in terms of the vertical position.

Two points require careful thought. First, different wind profiles may be considered \cite{BejiN}. We will later focus on the most common choice, a logarithmic wind profile, but for the time being the calculations are general. Second, matching the flow fields in water and air is a highly non-trivial matter, due to the presence of a turbulent layer between both media, which develops when the wind blows over the ondulated water surface. However, it as been shown that one may account for this in a phenomenological way by requiring that the air flow profile (\ref{airprofile}) match the water velocity not at the top of the water volume ($z=0)$, but at a height $z_0>0$. $z_0$ is known as the {\it roughness length}. For a wave of wavelength $2 \pi/k$ propagating with a phase velocity $c_0$ the roughness length has been shown to obey the so-called Charnock relation \cite{BejiN}
\begin{equation} \label{z0}
  z_0 = \frac{\Omega_{CH}}{k} \, \left(\frac{U_1}{c_0}\right)^2
  \ ,
\end{equation}
where $\Omega_{CH}$ is a phenomenological constant~\footnote{this constant is often denominated $\Omega$, for example in \cite{BejiN}. We therefore preserve the symbol, adding a subscript 'CH' in order to indicate that it is ultimately derived from the Chernock constant. However, $\Omega_{CH}$ is {\it not} related to the vorticity $\Omega$ in any way.}, the value of which is to be adapted to the specific wind profile and to the choice of the characteristic wind velocity $U_1$. 
In this approach, the air flow then vanishes at the top of the roughness layer $z_0$, rather than at the average position of the water-air interface ($z=0)$, i.e. we have
\begin{equation} \label{eq:U_at_z0}
  U(z_0)=0    
  \ .
\end{equation}
This is a widely used approximation, first proposed in Ref. \cite{Charnock}, and it is indeed appropriate for the ranges of wind speed considered here \cite{Fairall}. 

Note that the {\it empirical} expression (\ref{z0}) for the roughness length $z_0$ has been established in infinite deep water and on static water \cite{BejiN,Charnock}. Here we assume that this relation extends to our situation. 

In order to analyse the propagation of a wave at the water-air interface in our system we now establish the hydrodynamic equations in each domain, before coupling them via the appropriate boundary conditions.


\subsection{The water domain}
\label{water domain}

As a wave propagates on the water surface, for now without accounting for the effect of wind, the flow field is
\begin{equation}
\vec{U}_{w}(x,z,t)
=\big(U_{w}(x,z,t),0,W_{w}(x,z,t)\big)
= \left(U_0^{(\Omega)}(z) + u(x,z,t),0,w(x,z,t)\right)  
\end{equation}
in the reference frame of the water surface.
Here $U_0^{(\Omega)}(z)$ is the stationary flow field referred to in (\ref{profile}), to which perturbations are added as the wave propagates. 
This velocity field satisfies the Euler equations:
\begin{eqnarray}\label{Eulernonlinear}
U_{w,x} + W_{w,z} &=& 0 \\
U_{w,t} + U_w U_{w,x}+ W_w U_{w,z}&=&-\frac{1}{\rho_w}P_x\\
W_{w,t}+ U_w W_{w,x}+ W_w W_{w,z}&=&-\frac{1}{\rho_w}P_z-g
\ ,
\end{eqnarray}
where $P$ is the pressure, $g$ is the gravitational acceleration and $\rho_w$ the water density.
Subscripts to $U_w(x,z,t)$, $W_w(x,z,t)$ and $P(x,z,t)$ denote partial derivatives.

The boundary conditions to this flow are to be imposed at $z = \eta (x,t)$ and at $z = -h $. They are
\begin{eqnarray}
P&=& P_a,\quad \mbox{at}\quad z=\eta\label{EulerWaterBC1}
\\
\eta_t + U_w\eta_x -W_w &=& 0, \quad \mbox{at}\quad z=\eta \label{EulerWaterBC2} 
\\
W_w&=&0, \quad \mbox{at}\quad z=-h \ .
\label{EulerWaterBC3}
\end{eqnarray}
Here $P_a(x,z,t)$ is the variable air pressure, and $\eta_x$ and $\eta_t$ are partial derivatives of $\eta(x,t)$. Thus equation (\ref{EulerWaterBC1}) expresses the continuity of the pressure across the air/water interface. We introduce the {\it dynamic pressure} $\mathbf{P}(x,z,t)$, 
relative to the hydrostatic pressure in the undisturbed initial state, as
 \begin{equation}\label{reduced pressure}
 \mathbf{P}(x,z,t)=P(x,z,t) + \rho_w gz -P_{0} \ ,
\end{equation}
 where $P_0$ is a constant.
 
The perturbation to the free surface is $\eta(x,t)$, and $u(x,z,t)$, $w(x,z,t)$ as well as $\mathbf{P}(x,z,t)$ are those to the horizontal and vertical velocities as well as the pressure, respectively. Linearizing equations 
 (\ref{Eulernonlinear})-(\ref{EulerWaterBC3}) around the unperturbed state (\ref{profile}), 
 expressed in terms of $\mathbf{P}$ and the perturbations  $u$, $w$ and $\eta$, we have
 \begin{eqnarray}\label{Eulerlinear}
 u_x+w_z&=&0 \label{cont2}
 \\
 u_t + \Omega z \,u_x + \Omega \, w   &=&-\frac{1}{\rho_w}\mathbf{P}_x  \label{Euler_xlinear} 
 \\
 w_t + \Omega z \, w_x &=&-\frac{1}{\rho_w}\mathbf{P}_z  \label{Euler_zlinear}
 \ .
 \end{eqnarray}
The boundary conditions are also to be linearised, and they become 
\begin{eqnarray}
 \mathbf{P}(x,\eta,t) &=& P_a(x,\eta,t)+\rho_w g \eta -P_0 \label{EulerWaterlinearBC12}
 \\
\eta_t &=&w(0), \label{EulerWaterlinearBC22} 
\\
w(-h)&=&0
\label{EulerWaterlinearBC32}
\ .
\end{eqnarray}

\begin{figure}
\centering
\includegraphics[width=0.5\textwidth]{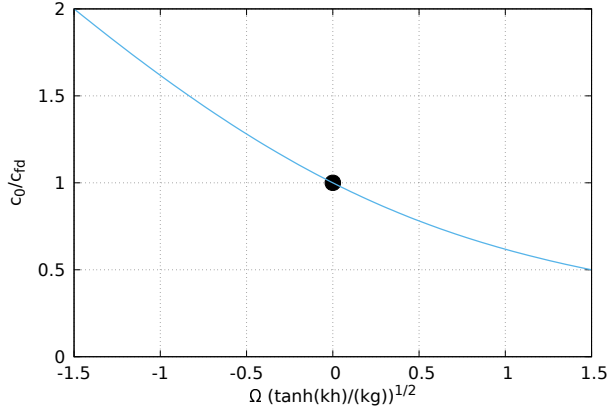}
\caption{ \label{theta-de-nu}
Effect of vorticity on a wave propagating on the water surface. The graph shows the multiplicative factor by which the phase velocity $c_0$ is modified as compared to the same system with no vorticity. The graph is based on Eq.~(\ref{+}), and vorticity is represented through a dimensionless expression  $\Omega \,\sqrt{\tanh(k h)} / \sqrt{k g}$, proportional to vorticity. By construction, the value $1$ is attained for vanishing vorticity, indicated by a circle. Negative vorticity increases the phase velocity, whereas positive vorticity slows the wave down. 
}
\end{figure}

The system of linear equations (\ref{cont2})-(\ref{EulerWaterlinearBC32}) can be solved assuming normal mode solutions  as
\begin{equation}\label{Fourier}
 \mathbf{P}=\mathcal{P}(z)\exp{(i\varphi)} ,\quad u=\mathcal{U}(z)\exp{(i\varphi)} , \quad
 w=\mathcal{W}(z)\exp{(i\varphi)} , \quad \eta=\eta_{0}\exp{(i\varphi)}
 \ ,
\end{equation}
with $\varphi= k(x-ct)$, where $k$ is the wavenumber, $c$ the phase speed and $\eta_0$ is a constant.
\\
Using equations  (\ref{Fourier}), (\ref{cont2}), (\ref{Euler_xlinear}), 
(\ref{Euler_zlinear}), (\ref{EulerWaterlinearBC12}) and  (\ref{EulerWaterlinearBC32}) we obtain 
\begin{eqnarray}
u(x,z,t)&=&2ai\exp{(-kh)}\exp{(i\varphi)}\cosh{k(z+h)}\label{linearu},\\
w(x,z,t)&=&2a\exp{(-kh)}\exp{(i\varphi)}\sinh{k(z+h),}\label{linearw} \\
\frac{1}{\rho_w} \mathbf{P}(x,z,t) &=&\frac{1}{\rho_w}[P_a(x,\eta,t)-P_0] + g\eta-2aic\exp{(-kh)}\exp{(i\varphi)} \big\{\cosh{(kh)} \label{linearpression}   \\
&&-\cosh{k(z+h)}\}+2ai\Omega\exp{(-kh)}\exp{(i\varphi)}\{-z\cosh{k(z+h)} \nonumber \\
&& - \frac{1}{k}\sinh{(kh)}+\frac{1}{k}\sinh{k(z+h)}\big\} \nonumber
\end{eqnarray} 
with some constant $a$. Now we calculate $(1/\rho_w)  \mathbf{P}_x$ from (\ref{linearpression}), and substituting into (\ref{Euler_xlinear}) we obtain
\begin{equation}\label{formula1}
ikg\eta+2a\exp{(-kh)}\exp{(i\varphi)}\big\{kc\cosh{(kh)}+\Omega\sinh{(kh)}\big\}=-\frac{P_{a,x}}{\rho_w}
\ .
\end{equation}
From (\ref{EulerWaterlinearBC22}) and (\ref{linearw}) we obtain
\begin{equation}\label{formula2}
-ik\eta c-2a\exp{(-kh)}\exp{(i\varphi)}\sinh(kh)=0
\ .
\end{equation}
Finally, it follows from (\ref{formula1}) and (\ref{formula2}), integrating over $x$, that
\begin{equation}\label{disprelation}
\eta \, \big\{g-kc^{2}\frac{\cosh{(kh)}}{\sinh{(kh)}}-c\Omega\big\}=\frac{-P_{a}+P_0}{\rho_w}
\ .
\end{equation}
For $P_{a}=P_0$, Equation~(\ref{disprelation}) leads to the very well known phase velocity $c=c_{0}$ 
\begin{eqnarray}\label{phasespeedc}
c_{0}&=&-\frac{\Omega}{2k}\tanh{(kh)}
\pm
\frac{1}{2}\sqrt{\frac{\Omega^2}{k^2}\tanh^{2}{(kh)}+\frac{4g}{k}\tanh{(kh)} }
\ ,
\end{eqnarray}
where we introduce the subscript '0' to refer to the reference state of a propagating wave on the water surface in the presence of shear flow, but without air flow.
\\

In order to clarify the discussion we choose to consider a wave propagating to the right in the remainder of the manuscript, which amounts to selecting the positive branch of Equation (\ref{phasespeedc}): 
\begin{eqnarray}\label{+}
c_{0}&=&-\frac{\Omega}{2k}\tanh{(kh)}
+\frac{1}{2}\sqrt{\frac{\Omega^2}{k^2}\tanh^{2}{(kh)}+\frac{4g}{k}\tanh{(kh)} }\\
&=& c_{fd}\left[ \sqrt{1+\frac{\Omega^2 \tanh{(kh)}}{4kg}} -\frac{\Omega\sqrt{\tanh{(kh)}}}{2\sqrt{kg}}\right] 
\ ,
\end{eqnarray}
where $c_{fd}$ is the finite depth velocity defined by $c_{fd}=(g\tanh{(kh)}/k)^{1/2}$. 
This phase velocity plays an important role when discussing the effect of wind, and we therefore illustrate its dependence on vorticity in Fig.~\ref{theta-de-nu} for later reference.

In the complete problem, however, in which we must account for the wind profile, the air pressure can no longer be taken to be constant: $P_{a}\neq P_0$.  In order to determine the phase velocity $c$ in the presence of wind, from (\ref{disprelation}), we thus need to determine the dynamic perturbation to the air pressure  $P_a$,  and its value at the interface position $z=\eta$ ultimately enters the balance equations at the water-air interface.


\subsection{The air domain}
\label{air domain}

Let us consider the linearized governing equation of a steady air
flow, with a prescribed mean horizontal velocity $U(z)$, as stated in (\ref{airprofile}). 
We are going to study perturbations to this mean flow in the $x$ and the $z$ components, where the subscript $a$ stands for \emph{air}: $u_a(x,z,t)$, $w_a(x,z,t)$, as well as the air pressure field $P_a(x,z,t)$, are the dynamic contributions as the wave propagates. $\rho_a$ is the air 
density, and we define 
$\mathbf{P}_a(x,z,t)=P_a(x,z,t) + \rho_a gz -P_{0}$.
In the reference frame where there is no surface flow
we have, from continuity and from the Navier Stokes equations:
\begin{eqnarray}
u_{a,x}+w_{a,z}&=&0 \label{continuityair}\\
u_{a,t} + U(z)u_{a,x} + U_{z}(z)w_a&=&-\frac{1}{ \rho_a}\mathbf{P}_{a,x} 
\label{uair} \\
w_{a,t} + U(z)w_{a,x} &=&-\frac{1}{ \rho_a }\mathbf{P}_{a,z}\label{wair}
\ .
\end{eqnarray}
To these equations governing the flow field we must add the appropriate boundary conditions. The first one is the kinematic boundary condition
\begin{equation}
 \eta_t + U(z_0)\eta_x = w_a(z_0) \label{etaair}
 \ .
 \end{equation}
 As discussed above, it is to be evaluated at the aerodynamic sea surface roughness $z_0$. Exploiting the fact that the wind flow vanishes at the roughness height, Eq.~(\ref{eq:U_at_z0}), the kinetic boundary condition (\ref{etaair}) finally reduces to 
\begin{equation}\label{etaairreduced}
\eta_t = w_a(z_0)
\ .
\end{equation}

Next we introduce normal mode expressions, as in (\ref{Fourier}), also for the air flow,
\begin{equation}
\mathbf{P}_a=\mathcal{P}_a(z)\exp{(i\varphi)} \ , \ u_a=\mathcal{U}_a(z)\exp{(i\varphi)} \ , \ w_a=\mathcal{W}_a(z)\exp{(i\varphi)} \ ,
\end{equation}
and we add the following 
boundary conditions on $\mathcal{W}_a$ and $\mathcal{P}_a $:
\begin{eqnarray}
\lim\limits_{ z \to +\infty}(\mathcal{W}_a' + k\mathcal{W}_a) &=& 0 \label{W_ainfinite}\\
\lim\limits_{ z \to z_0} \mathcal{W}_a &=& W_{0}
\label{W_ainz0}\\
\lim\limits_{ z \to+\infty} \mathcal{P}_a &=& 0
\ . \label{P_ainfinite}
\end{eqnarray}
The first condition imposes that, far from the interface, the perturbation of air flow must decrease exponentially. The remaining conditions set the vertical component of the wind speed in terms of the wave motion at the sea surface, and require pressure continuity across the water-air interface.

Finally, using equations (\ref{continuityair})-(\ref{wair}) and (\ref{P_ainfinite}) we have
\begin{eqnarray}
 w_a(x,z,t)&=&\mathcal{W}_a\exp{(i\varphi)}
 \ \label{w_a}\\
 u_a(x,z,t)&=&\frac{i}{k}\mathcal{W}_{a,z}\exp{(i\varphi)}
 \ \label{u_a}\\
 \mathbf{P}_a(x,z,t)&=&ik\rho_a\exp{(i\varphi)}\int^{\infty}_{z} \big[U(z')-c\big]
 \mathcal{W}_{a}(z')dz'
 \ .\label{mathP_a}
\end{eqnarray}

By eliminating the pressure from the Euler equations we obtain the well-known Rayleigh equation  \citep{Rayleigh}, which holds $\forall z~ \backslash~ z_{0} < z < +\infty$ 
\begin{equation}\label{Rayleigh}
 \big[U(z) - c\big] \, (\mathcal{W}_{a,zz} - k^2 \mathcal{W}_a) - V_{zz} \mathcal{W}_a = 0
\ .\end{equation}
This is also known as the inviscid Orr-Sommerfeld equation. It is singular at the 'critical' or 'matched' height $z_{c} = z_0 e^{c \kappa/ u_{*}} > z_{0} > 0$, where $U(z_{c}) = c$. 
We recall that this model assumes any eddies or other non-linear phenomena to be accounted for by the roughness height $z_0$, and therefore the bulk of the air flow is non-turbulent.

\subsection{Matching the flow at the interface}

Recall that, in equations (\ref{w_a})-(\ref{Rayleigh}), neither the function determining the perturbation to the air flow $\mathcal{W}_a(z)$ nor the phase velocity $c$ are known: indeed, finding the phase velocity in the presence of wind, as well as the shear flow in water, is precisely the object of our calculation. 
This is achieved by applying the appropriate boundary conditions across at interface. In order to do so we have to determine the air pressure field $P_a(x,\eta,t)$. 
We obtain
\begin{equation}\label{integralforP}
P_a(x,\eta,t)=P_0 - \rho_a g \eta 
+ {ik\rho_a\exp{(i\varphi)}\int^{\infty}_{z_0} \big[U(z)-c\big] \mathcal{W}_a(z) dz}
\ ,\nonumber
\end{equation}
where the lower bound for integration may be taken to be the constant roughness height $z_0$ instead of $z=\eta$, since we are studying the linear problem.

Finally, using equation (\ref{etaairreduced}) to eliminate the term $ik\rho_a\exp(i\varphi)$ and substituting $P_a$ given by equation (\ref{integralforP}) into (\ref{disprelation}) we obtain what is effectively the dispersion relation of the problem, as it fixes the phase velocity $c$:
\begin{equation}\label{equationforc}
g(1-\epsilon) + c\left\{\frac{s k^2}{W_0}I_1-\Omega\right\}-c^2\left\{\frac{sk^2}{W_0}I_2 + k\coth(kh)\right\}=0
\ .
\end{equation}
Here $s=\rho_a/\rho_w$ and the integrals $I_1 $ and $I_2$ are defined as
\begin{equation}\label{I1andI2}
I_1=\int_{z_0}^{\infty} U\mathcal{W}_a dz,\quad I_2=\int_{z_0}^{\infty}\mathcal{W}_a dz
\ .
\end{equation}
The density of air being small compared to that of water, relation (\ref{equationforc}) may be expanded as a series in terms of $s=\rho_a/\rho_w \sim 10^{-3}$:
\begin{equation}
c=c_0 + s c_1 + O(s^2)
\ .
\end{equation}
An explicit expression for the first order term $c_1$ can be established using a perturbation method, which consists in solving the Rayleigh equation (\ref{Rayleigh}) based on the zero-order phase velocity $c_0$. This approach, which follows \cite{BejiN}, is pursued in the next section.

The dispersion relation can be deduced from (\ref{equationforc}), once we have evaluated the integrals $I_1$ and $I_2$ given by  (\ref{I1andI2}): once it is known, its imaginary part carries the information on wave growth. For this, in turn, the profile $\mathcal{W}_a(z)$ is to be established as a solution to the Rayleigh equation (\ref{Rayleigh}). Before doing this numerically, we pursue to introduce the growth coefficients we intend to analyse.


\subsection{Wave growth during propagation in the presence of vorticity and wind}
\label{wave growth}

Once the function $\mathcal{W}_a(z)$ is known, its imaginary part intervenes in the dispersion relation (\ref{equationforc}), and sets the complex part to the phase velocity $c$. This directly yields the growth rate $\gamma$ of the wave amplitude $\eta(x,t)$, defined as 
\begin{equation}
 \gamma =k\Im{(c)}
 \ ,
\end{equation}
where $\Im$ stands for the imaginary part.

The theoretical and numerical results for the growth rate $\gamma$ are traditionally studied and computed in terms of dimensionless parameters. To the established parameters $\delta$, $\theta_{dw}$ (see \citet{YoungVerhagen1}, \citet{YoungVerhagen2} and \citet{Manna1}) we now add a third parameter $\nu$ characterising vorticity. These are defined as
\begin{equation}\label{delta}
\delta = \frac{gh}{U^2_1},\quad \theta_{dw}=\frac{1}{U_1}\sqrt{\frac{g}{k}},\quad \nu=\Omega\frac{U_1}{g}
\end{equation}

It is thus immediately clear that, in contrast to deep water, where the growth rate can be characterised by as a single graph showing the growth rate as a function of wave age, the situation is more complex here: the presence of two additional dimensionless parameters imply that we will instead be dealing with a \textit{two-parameter family of curves}.

Note that the parameters defined in (\ref{delta}) have direct physical meaning. 
The (square root of) dimensionless parameter $\delta$ relates the phase velocity of a shallow water wave to the wind velocity. The parameter $\theta_{dw}$ is the analogous ratio of the deep-water phase velocity and the wind speed. Finally,  $\nu$ is a measure for the importance of the velocity gradient due to vorticity.

The parameter $\theta_{dw}$ is often referred to as the 'wave age', or at least a theoretical equivalent of this observational quantity, and is therefore important for making contact with field observations and literature.
It has been put forward recently \cite{Manna1} that this notion can be generalized to water of finite depth, by defining the {\it finite depth wave age} $\theta_{fd}$ as
\begin{equation}\label{waveagefinite}
\theta_{fd}=\frac{1}{U_1}\sqrt{\frac{g}{k}}\sqrt{\tanh({kh})}=
\theta_{dw} T^{1/2}
\ ,\end{equation}
where 
\begin{equation}\label{tanh}
 T = \tanh(\frac{\delta}{\theta_{dw}^2})
 \ .
\end{equation} 
Expression (\ref{waveagefinite}) is a depth dependent parameter, which is constructed to interpolate between deep water and shallow water situations: indeed, for a finite and constant
$\theta_{dw}$, we have $\theta_{fd}\sim \theta_{dw}$ when  $\delta \rightarrow \infty$ and
 $\theta_{fd}\sim \delta^{1/2}=\sqrt{gh}/U_1$ if $\delta \rightarrow 0$. Therefore $\theta_{fd}$ reduces to the ratio of the appropriate phase velocity and the wind speed, in both limiting cases. 
 In the following section we will extend this concept to flow with constant vorticity.
 \\
 
We can now proceed to determine the dispersion relation and deduce the parameters characterising wave growth.
The following  non-dimensional variables and scalings will be used in order to non-dimensionalise the growth coefficients:
\begin{equation}\label{parameters}
 U=U_1\hat{U},\quad \mathcal{W}_a=W_0\hat{\mathcal{W}_a},\quad
z=\frac{\hat{z}}{k},\quad
c = U_{1} \hat{c},\quad t = \frac{U_{1}}{g} \hat{t}
\ ,
\end{equation}
 where 'hats' indicate dimensionless quantities.

Using (\ref{delta}) and (\ref{parameters}) in equations (\ref{phasespeedc}) and  (\ref{equationforc}), and retaining only first order terms in $s$, we obtain expressions for $\hat{c}_0$ and $\hat{c}$.
The zero-order terms reproduce Eq.~(\ref{+}), as expected. The linear term yields
\begin{equation}
\label{c-hat-c}
  \hat{c}
  = \hat{c}_0 + s \,   \left(-\frac{\hat{c}_0}{2-\nu\hat{c}_0}+\frac{\hat{c}_{0}^{2}}
  {\theta_{dw}^2[2-\nu\hat{c}_0]}[\hat{I}_{1}-\hat{c}_0\hat{I}_{2}] \right)  
  \ .
\end{equation}
which is the desired dispersion relation under the effect of wind.
\\

Finally the growth of the amplitude $\eta$ with time is given by
\begin{equation}
\exp^{\gamma t}=\exp^{\hat{\gamma}\hat{t}}
\end{equation}
with the corresponding growth rate
\begin{equation}\label{eq:gamma}
 \hat{\gamma}=s\frac{\Im{(\hat{c}_1})}{\theta_{dw}^2}=s \frac{\hat{c}_0^{2}}{\theta_{dw}^4[2-\nu\hat{c_{0}}]}\left(\Im(\hat{I}_1)-
 \hat{c_{0}}\Im{(\hat{I}}_2)\right)
\ .
\end{equation}
Hence we have established an expression for the growth rate $\hat{\gamma}$ for a given physical situation, i.e. a given set of parameters $(\delta,\theta_{dw},\nu)$. 

Our results also  explore the $\beta$-Miles parameter, related by
\begin{equation}\label{betaMiles1}
  \beta=\frac{2\hat{\gamma}}{s}\theta_{dw}^{2}\hat{c}_0,
\end{equation}
where we have taken $\beta$ as it is usually defined, via 
$\Im(c) =  (s/2)\beta c_0 (\frac{U_{1}}{c_{0}})^{2}$.

Using (\ref{gamma}) in (\ref{betaMiles1}) we obtain
\begin{equation}\label{betaMiles2}
\beta=2\frac{\hat{c}_0^3}{\theta_{dw}^2[2-\nu\hat{c}_0]}\left(\Im(\hat{I}_1)-
 \hat{c}_0\Im{(\hat{I}}_2)\right).
\end{equation}

Finally, another important parameter to be considered is the dimensionless fractional energy increase per radian, defined as
\begin{equation}
\hat{\Gamma}=\frac{c_{0,g}}{\omega_0E}\frac{dE}{dx}=2\gamma\frac{c_{0,g}}{\omega_0c_0}
\ ,
\end{equation}
with $c_{0,g}$, $c_{0}$ the group and  phase velocity of the dominant (spectral peak) waves of frequency $\omega_0$.
For $\Omega=0$ the ration $c_{0,g}/c_{0}$ is given by the very well known expression
\begin{equation}\label{Cdivc}
\frac{c_{0,g}}{c_{0}}=\frac{1}{2}\left[1+ \frac{2kh}{\sinh{(2kh)}}\right],\quad \mbox{for}\quad \Omega=0.
\end{equation}
Here we are interested in the generalization of (\ref{Cdivc}) to the presence of vorticity, which is given by
\begin{equation}\label{CdivcOmega}
\frac{c_{0,g}}{c_{0}}
=
\frac{1}{2}\left( \left[1+\frac{2kh}{\sinh{(2kh)}}\right]+\left[1-\frac{2kh}{\sinh{(2kh)}}\right]
\times 
\sqrt{\frac{\Omega^2 T}{\Omega^2 T+4gk}} \, \right)
\ ,
\end{equation}
for all $\Omega$. From this equation (\ref{CdivcOmega}) yields  $\hat{\Gamma}$, which reads 
\begin{equation} \label{eq:Gamma}
\hat{\Gamma}
=
\frac{\theta_{dw}}{T^{1/2}}\hat{\gamma}\left( \left[ 1+\frac{2\delta}{\theta_{dw}^2\sinh{(\frac{2\delta}{\theta_{dw}^2})}}\right]+ 
\left[1-\frac{2\delta}{\theta_{dw}^2\sinh{(\frac{2\delta}{\theta_{dw}^2})}}\right]
\times 
\sqrt{ \frac{\nu^2\theta_{fd}^2}{4+\nu^2\theta_{fd}^2} } \,
\right) 
\,
\frac{1}{-\frac{\nu\theta_{fd}}{2} + {\sqrt{1+\left(\frac{\nu \theta_{fd}}{2}\right)^2}}}
\ ,
\end{equation}
which is the dimensionless fractional energy increase per radian, accounting for the effect of vorticity.


\section{Results and discussion} \label{numerical}

In this section we exploit the results we have obtained above, based on the analytical expressions but aided by a numerical approach when required. To this end we follow the approach by Beji and Nadaoka \cite{BejiN}, which consists in numerically solving the Rayleigh equation in order to determine the velocity profile ${\mathcal{W}_a(z)}$ in the air domain. This requires handling the singularity of the Rayleigh equation (\ref{Rayleigh}), which is done by establishing an analytical approximation valid close to the singularity, i.e. around the height $z_c$ where the wind velocity equals the phase velocity. From this, starting values are deduced which serve to initialise a solver for differential equations.

At this stage we now focus our analysis on a specific air flow field $U(z)$, which is the logarithmic wind profile with 
\begin{equation}\label{logprofile}
 U(z)=U_{1} \ln(\frac{z}{z_0})
 \ .
\end{equation}
This relation is commonly used to describe the vertical dependency of the horizontal mean wind speed \citep{Garratt}. This can be justified based on scaling arguments and solution matching between the near-surface air layer and the geostrophic air layer (see \citet{Tennekes}).

\subsection{Accounting for vorticity in the wave age}

The first point we address is the notion of a generalised wave age, which is the parameter $\hat{c}_0$ already defined via (\ref{parameters}) as the ratio
\begin{equation}\label{waveage}
\theta = \frac{c_0}{U_1} = \hat{c}_0
\ ,
\end{equation}
with the phase velocity $c_0$ given by Eq.~(\ref{+}).
To see how this extends the existing definition, recall that the notion of wave age was defined originally, for deep water, as the ratio $c_{0,dw}/U_1$. It thus measures the phase velocity of a wave propagating {\it without wind} as compared to the wind speed.
A similar parameter $\hat{c}_{0,fd} = c_{0,fd}/U_1$ has been introduced in \cite{Manna1} as the generalisation of the wave age to finite water depth. Setting $\theta_{fd} = c_{0,fd}/U_1$, this characterises wave propagation on the surface of a finite depth water body. No vorticity was considered at this stage.
In our analysis, the expression for the phase velocity as established in Eq.~(\ref{+}) now suggests that the parameter defined in Eq.~(\ref{waveage}) directly generalises the notion of wave age further, by accounting for the presence of vorticity.

With this definition, the wave age in the presence of vorticity is deduced directly from Eq.~(\ref{+}) as
\begin{equation}
\label{theta-general}
\theta 
= c_0/U_1 = 
\theta_{fd} \, \left( -\frac{\nu\theta_{fd}}{2}+ \sqrt{1+\left(\frac{\nu\theta_{fd}}{2}\right)^2}  \right)
\ .
\end{equation}

As an aside, it may be worth pointing out that the choice of reference is important here: the ratio (\ref{waveage}) is to be taken with velocities expressed in the reference frame of the water surface, which we have already adopted here.
Defining the wave age in this way is only meaningful with respect to this particular reference frame. 
This is because the variable $U_1$ defines the typical wind scale in the reference frame of the water surface. It cannot simply be transformed to a different frame, as transforming  constant velocity to the wind profile ($\ref{logprofile}$) adds an overall constant, which is not equivalent to modifying $U_1$. Rather, the definition of $U_1$ is intrinsically based on the wind speed {\it relative to the water surface}, as this is what the Charnock relation (\ref{z0}) matches the wind profile to, and therefore $U_1$. Note that this consideration is new to the system combining waves with both wind and surface flow.

All other physical quantities concerning the growth of waves ($\gamma$ and $\beta$ for instance) remain unchanged through a Galilean transformation from one reference frame to another. Rather expectedly, calculations in the earth-bound frame  are found to lead to the identical results when the relative velocity between reference frames is added to the phase velocity.

\subsection{Miles growth parameter $\beta$}

\begin{figure}
    \centering
    \includegraphics[width=0.7\textwidth]{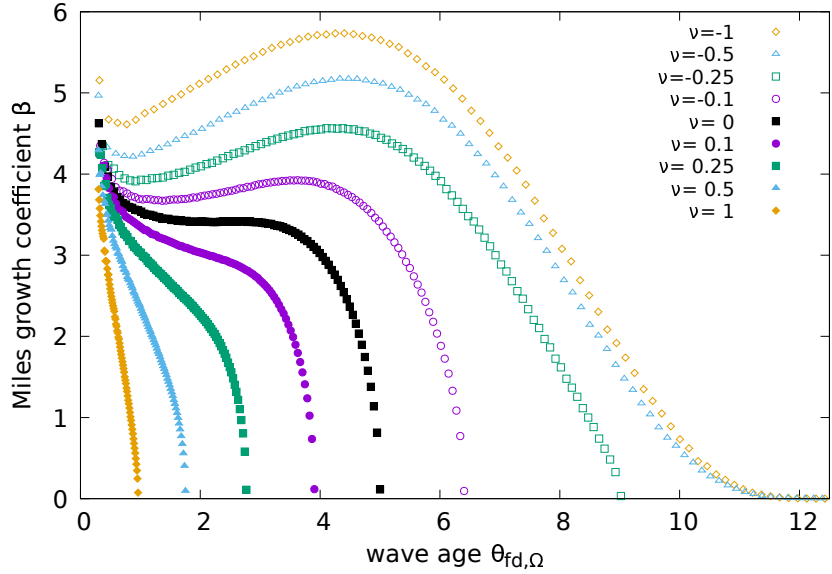}
    \caption{\label{beta-1}
    The Miles' growth coefficient $\beta$, as defined in (\ref{betaMiles2}) and obtained by numerically solving the Rayleigh equation (\ref{Rayleigh}). All plots are for the same depth parameter $\delta=25$. Each line corresponds to a given value of vorticity, according to the parameter $\nu$ indicated in the legend. The black line, for $\nu=0$, reproduces the known result for wave growth in finite depth \cite{Manna1}. Negative vorticity increases the growth coefficient, whereas positive vorticity reduces it. 
    }
\end{figure}

We first discuss the effect of vorticity on the Miles growth parameter $\beta$. In Fig.~\ref{beta-1}, results are shown for the Miles parameter
$\beta$ as a function of the vorticity-corrected wave age $\theta_{fd,\Omega}$.
All graphs are  calculated for the same depth parameter $\delta=25$, but correspond to different values of the vorticity parameter $\nu$. In all our plots the black curve corresponds to zero vorticity, for easy reference, and therefore the black graph of  Fig.~\ref{beta-1} reproduces the result of  Fig. 2 in reference \cite{Manna1}.

The most remarkable feature of $\beta(\theta)$ in the absence of vorticity is the steep drop to zero as the wave age approaches $\sqrt{\delta}=5$. This has been discussed in \cite{Manna1}, and implies a maximal wave age in shallow water: it is directly related to the fact that there is a maximum phase velocity, which a propagating wave cannot exceed in finite water depth. In Fig.~\ref{beta-1} it is clear that this maximum wave age is modified by vorticity: it is increased for negative values of $\nu$, whereas it is reduced for positive values. 

\begin{figure} 
    \centering
    \includegraphics[width=0.6\textwidth]{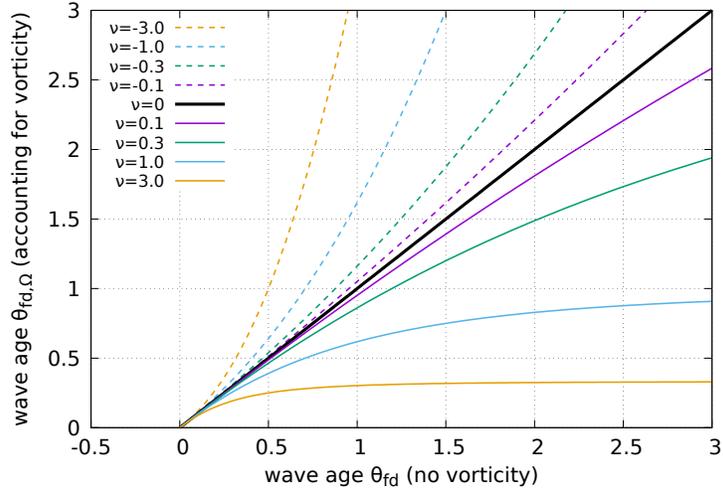}
    \caption{ \label{theta_vs_thetafd}
    The generalised wave age $\theta_{fd,\Omega}$, which incorporates the effect of vorticity, can be written as a function of $\theta_{fd}=\theta_{fd,\Omega=0}$, the wave age without vorticity. 
    The mathematical relation is given by Eq.~(\ref{theta-general}). Colours represent different values of vorticity; solid lines correspond to positive vorticity  whereas dashed lines of the same colour correspond to negative vorticity of the same negative values. Note that only in infinite depth all wave ages are accessible: in a finite depth system the wave age  has an upper bound of $\theta_{fd}^{(max)} =\sqrt{\delta}$. This entails a maximum value $\theta_{fd,\Omega}^{(max)}$ also in the presence of vorticity, but this maximum accessible wave age is {\it reduced} for positive vorticity and {\it increased}  for negative vorticity.
    }
\end{figure}

One way of visualising this is to plot the generalised wave age $\theta_{fd,\Omega}$, which accounts for vorticity, as a function of the zero-vorticity wave age $\theta_{fd}$. This is shown in Fig.~\ref{theta_vs_thetafd}. The dashed black line indicates that $\theta_{fd,\Omega}$ and $\theta_{fd}$ are identical for zero vorticity. The effect of vorticity is then quite different according to its sign. To discuss this, we first return to the case of zero vorticity, and recall that the upper bound to the wave ages which are accessible in water of finite depth (see \cite{Manna1}) is
\begin{equation}\label{thetafdmax}
    \theta_{fd} \le \theta_{fd}^{(max)}=\sqrt{\delta} =\frac{\sqrt{gh}}{U_1}
\ .
\end{equation}

Consequently, the maximum possible wave age decreases with depth, and it is only for infinitely deep water that a potentially unlimited domain of accessible wave ages is to be considered.

This is no longer the case in the presence of positive vorticity. 
Indeed, for $\nu>0$ the graph of the vorticity-corrected wave age asymptotically attains a plateau, the value of which is easily seen to be $1/\nu$ in deep water or for strong vorticity. Therefore one impact of positive vorticity is to set a maximum wave age even for a deep water wave.
For negative vorticity, no such upper bound is set in deep water;
instead, the vorticity correction increases the wage age.

\begin{figure}
    \centering
    \includegraphics[width=0.7\textwidth]{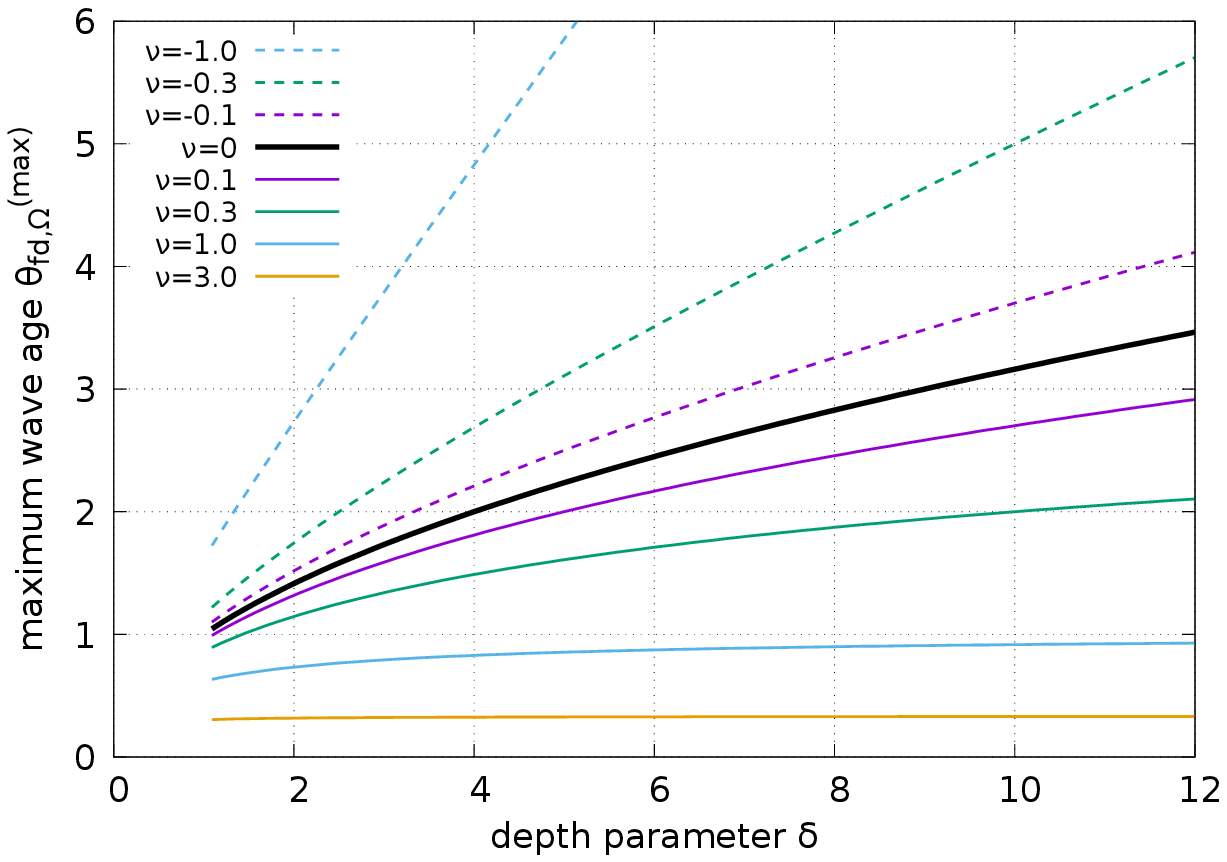}
    \caption{ \label{theta-max}
    The maximum achievable value for the wave age 
    (Eq.~(\ref{thetacritique})) as a function of the depth parameter $\delta$, for selected vorticity values of both signs. For a static sea ($\nu=0$, black curve) we have $\theta_{fd}^{(max)} = \sqrt{\delta}$, as is known from \cite{Manna1}. For positive vorticity, the maximum value is reduced; in turn, it is increased for negative vorticity. 
    }
\end{figure}

To complete the argument, we state the expression of the upper bound in the velocity-corrected wave age. Using Eq.~(\ref{theta-general}) and (\ref{thetafdmax}) we obtain, 
\begin{equation}\label{thetacritique}
 \theta_{fd,\Omega}^{(max)}
 =
 \theta_{fd}^{(max)}
 \left[
  -\frac{\nu\theta_{fd,c}}{2}
  +
  \sqrt{ 1+\frac{\nu^2\theta_{fd,c}^2}{4}} \,
  \right]   
  =
  \sqrt{\delta}
 \, \left[-\frac{\nu\sqrt{\delta}}{2}+ \sqrt{1+\frac{\nu^2\delta}{4} } \, \right] 
\ .
\end{equation}
Note that this correctly reduces to $\delta^{1/2}$ for $\nu=0$, as it must.  From this expression it follows directly that negative vorticity increases the maximum wave age, and hence the maximum wave propagation speed.  Therefore, negative velocity produces older seas while positive vorticity has the opposite effect. The corresponding graph is shown in Fig.~\ref{theta-max}.

\begin{figure}
    \centering
    \subfloat[$\delta=4$]{
        \includegraphics[width=0.7\textwidth]{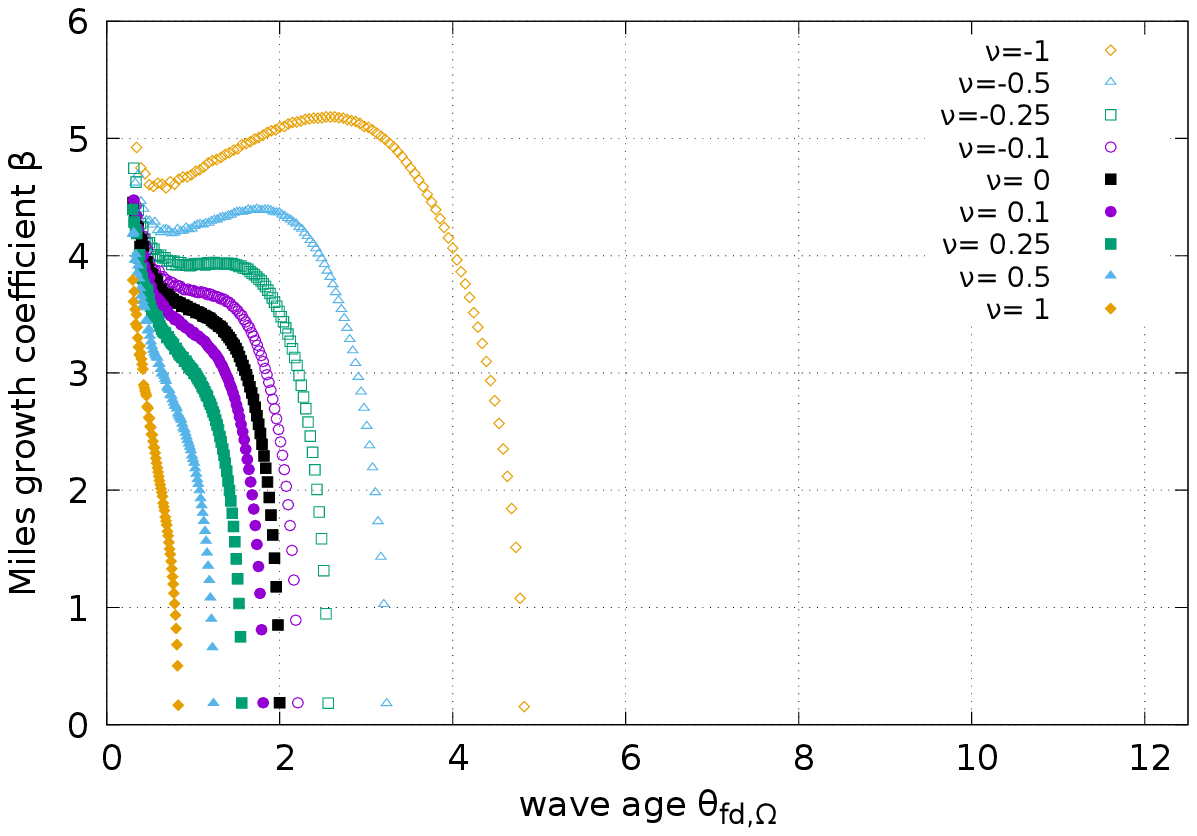}}
        \\
   \subfloat[$\delta=81$]{
        \includegraphics[width=0.7\textwidth]{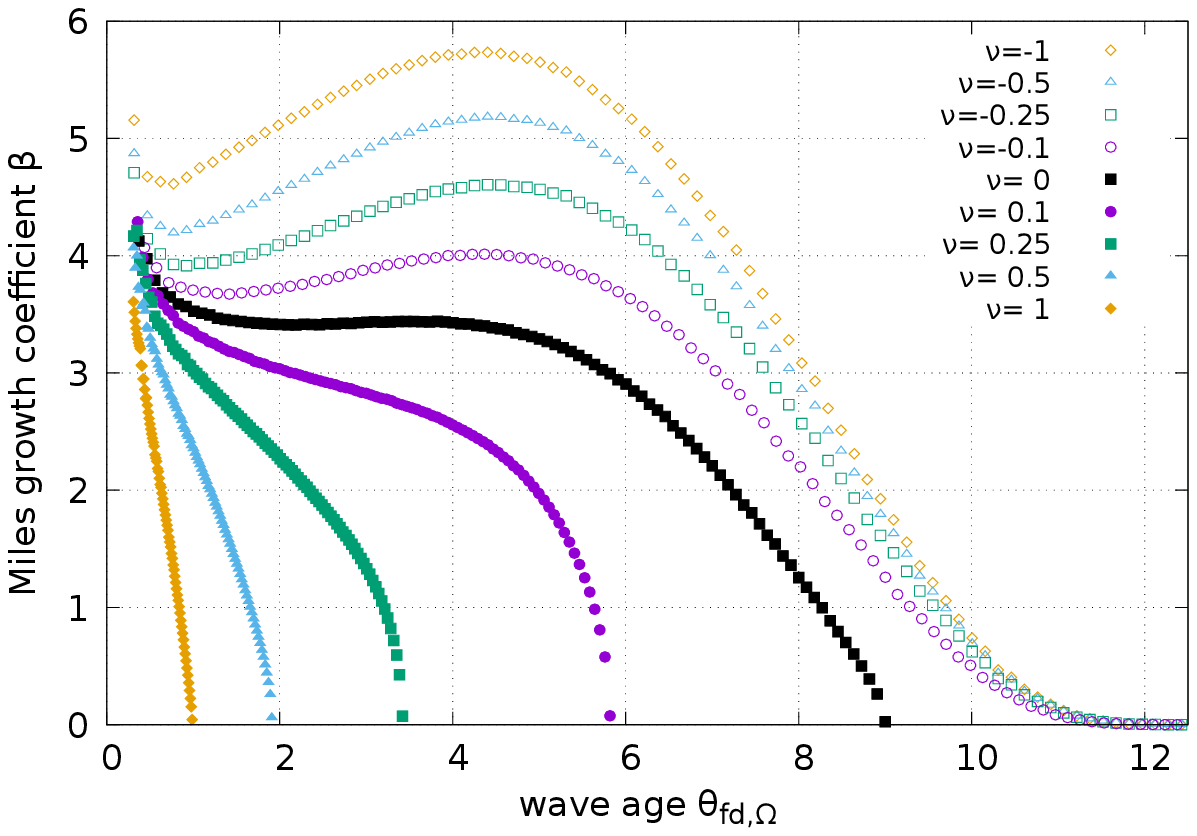}}
    \caption{\label{beta-2}
    The Miles growth parameter $\beta$ as a function of the wave age $\theta_{fd,\Omega}$, as in Fig.~\ref{beta-1} but for two additional values of the depth parameter $\delta$ in order to show the effect of the depth parameter.
    }
\end{figure}

To complete the discussion of the Miles growth coefficient $\beta$, we now contrast Fig.~\ref{beta-1} by 
similar plots in Fig.~\ref{beta-2}, obtained for two different values of $\delta$. The values used for the vorticity parameter $\nu$ are identical in all three graphs. We observe the  same tendencies, with a major difference worth pointing out. Indeed, the vorticity parameter $\nu$ has a weaker effect at small depth: curves for identically spaced values of $\nu$ are closer one to another for $\delta=4$. In comparison, the effect of vorticity is larger in greater depth, exemplified by $\delta=81$. 

As concerns practical implications, however, one must keep in mind that an identical vorticity parameter does not mean an identical current at the water surface. Rather, it is the combination of vorticity and depth which sets the shear current, through the difference of surface and bottom currents. Indeed, non-dimensionalising this current by the wind speed $U_1$ we have $(U_s-U_b)/U_1=\Omega h/U_1=\nu\delta$, and thus both $\nu$ and $\delta$ intervene.

\begin{figure}
    \centering
   \parbox{\textwidth}{
   \includegraphics[width=0.7\textwidth]{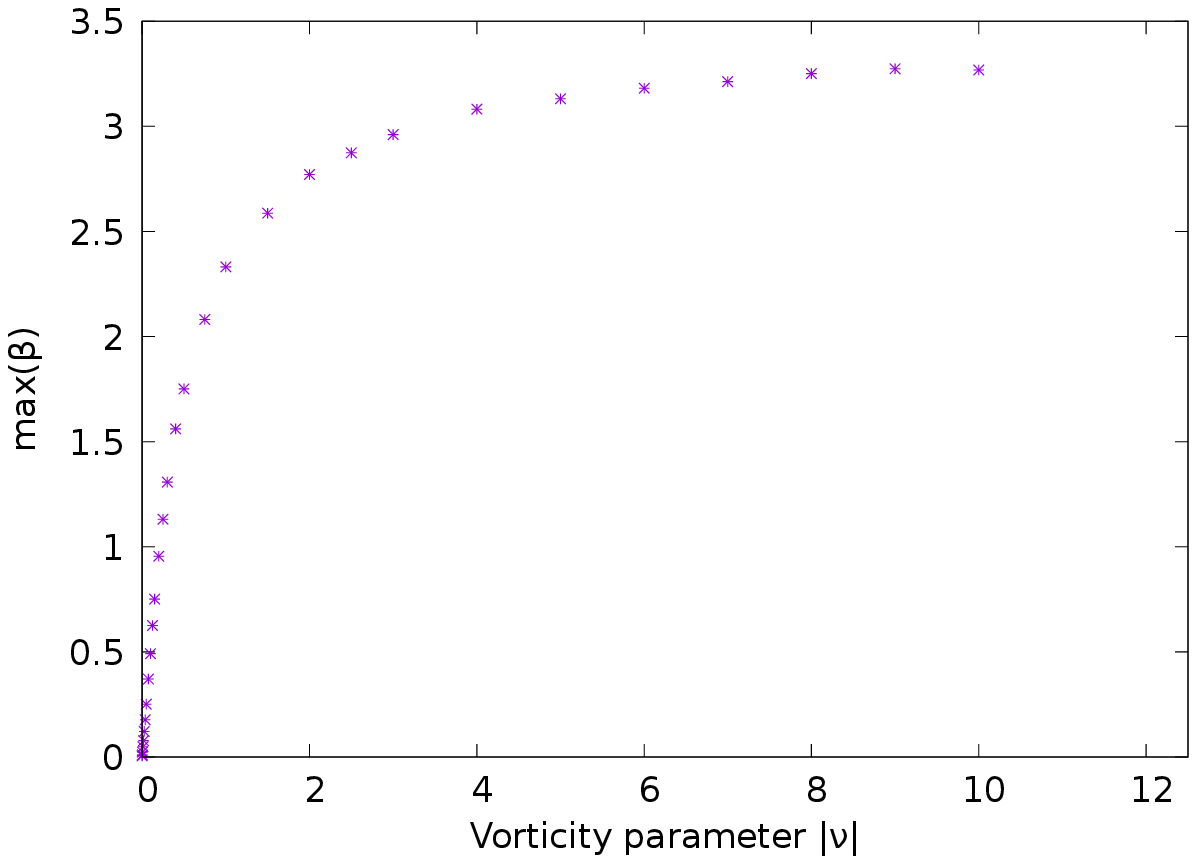}
    \put(-220,45){\includegraphics[width=0.4\textwidth]{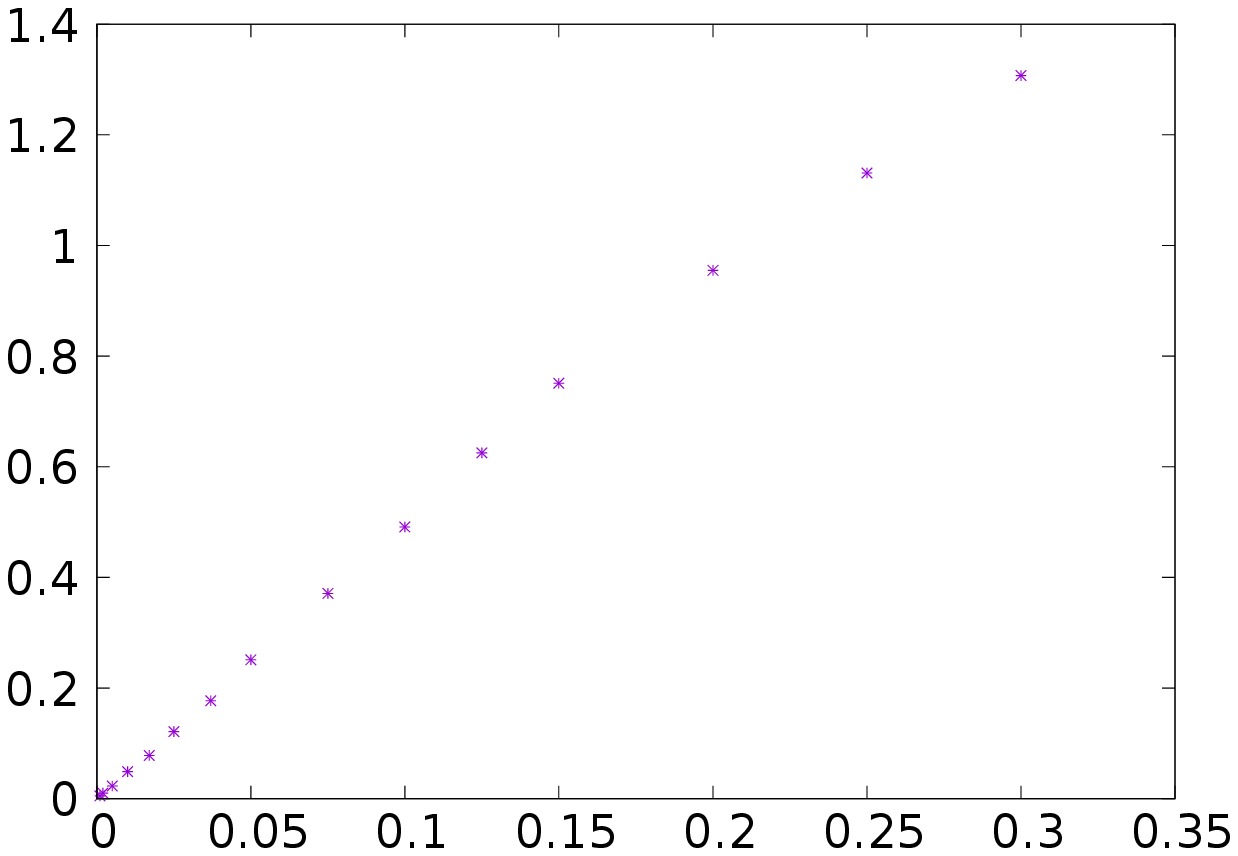}}
    }
    \caption{\label{max-beta}
    The maximum value of Miles' growth coefficient $\beta$ as a function of  (negative) vorticity $\nu$. We report the difference between the maximum and the plateau value at zero vorticity. Negative vorticity increases the Miles coefficient, until it saturates. The inset shows a zoom on the same data, for small values of vorticity, indicating that the dependency is initially linear.
    }
\end{figure}

Apart from this important difference, all families of curves for a given value of $\delta$ show the same behavior, and the same enhancement for negative vorticity. More specifically, a maximum is observed for $\beta$ as soon as negative vorticity is present. This maximum increases with increasing (negative) vorticity, a novel feature: changes in water depth cannot provoke this effect.
In Fig.~\ref{max-beta} we analyse the shift in $\beta_{max}$ with respect to the plateau value at zero vorticity, plotting it as a function of $\nu$.
The shift is clearly linear for (negative) vorticities up to $\nu \leq -0.5$, as shown in the inset of Fig.~\ref{max-beta}. This is  followed by $\beta$ saturating at around $\nu\approx2$ at a value of $\beta_{max}\approx6.5$.
The implication is that there is an upper bound to the increase of $\beta$ which can be provoked by vorticity.

\begin{figure}[h]   
    \begin{center}
    \parbox{0.6\textwidth}{
      \subfloat[$\delta=25$]{
        \includegraphics[width=0.6\textwidth]{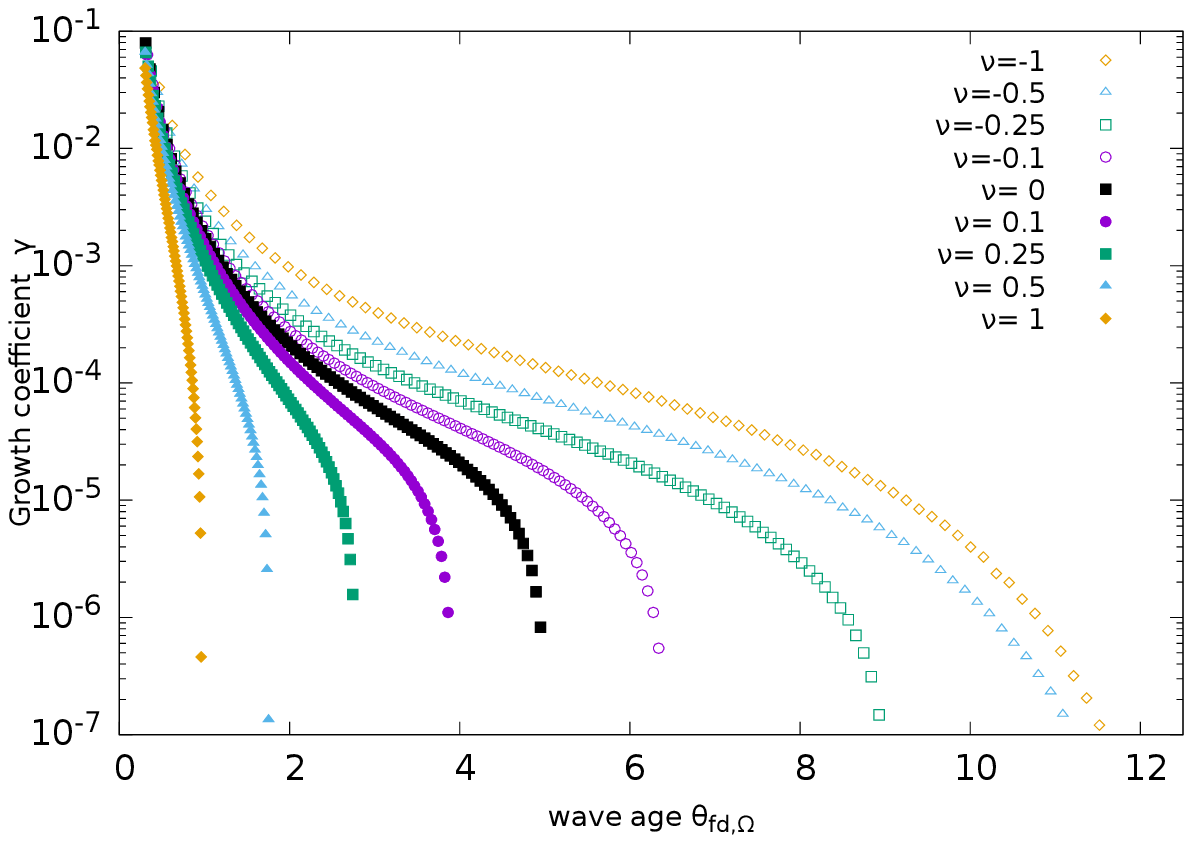}}
        \label{gamma-delta=25}
      } 
    \parbox[t]{0.35\textwidth}{
    \subfloat[$\delta=4$]{
      \includegraphics[width=0.35\textwidth]{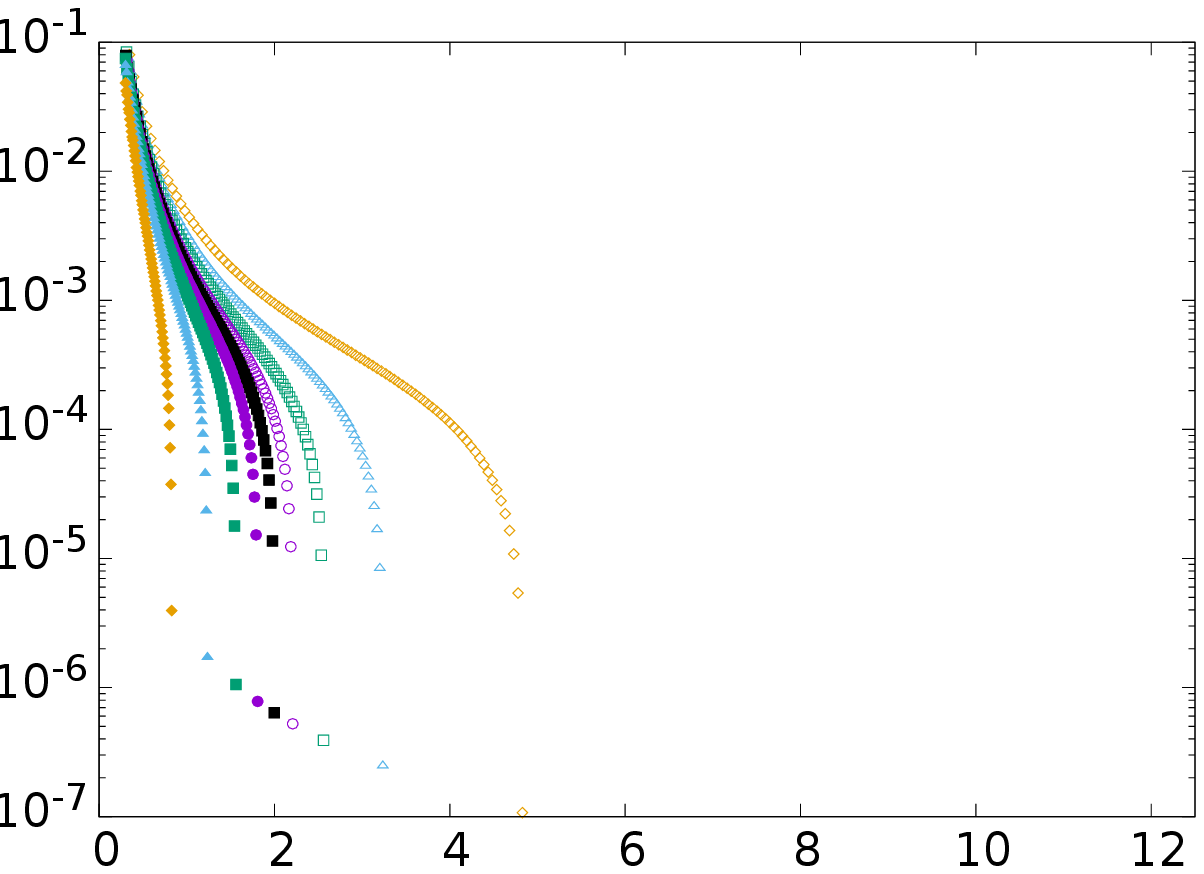}
      \label{gamma-delta=4}
      }
      
     \subfloat[$\delta=81$]{
       \includegraphics[width=0.35\textwidth]{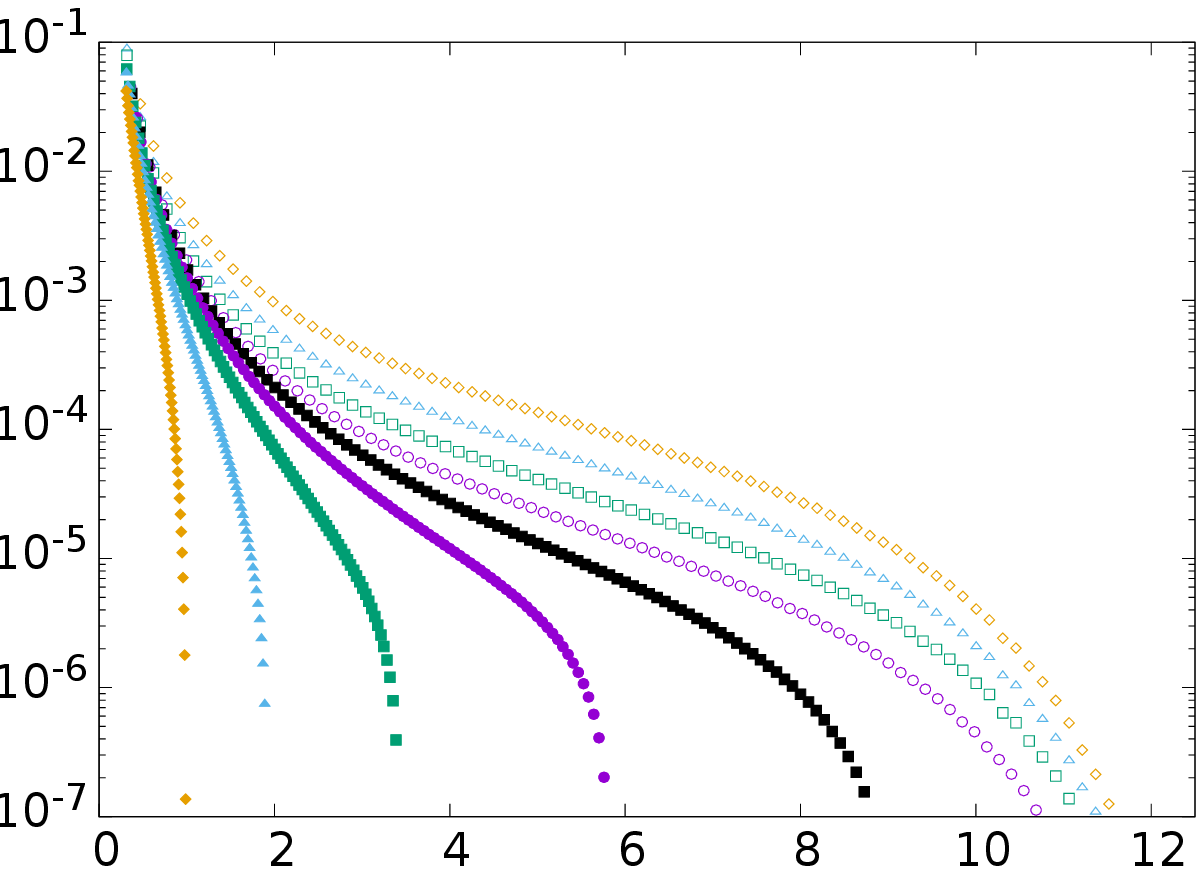}
       \label{gamma-delta=81}
      }
    }
    \caption{The amplitude growth coefficient $\hat{\gamma}$, defined by Eq.~(\ref{eq:gamma}), as a function of the wave age $\theta$ for three values of the depth parameter. In (b) and (c) the data ranges, as well as symbols, 
    are identical to those indicated in (a). }
    \label{gamma}
  \end{center}
\end{figure}

The increase of $\beta$ is confirmed by the evolution of the amplitude growth coefficient $\hat{\gamma}$ with vorticity. In Fig.\ref{gamma} we have plotted three families of curves, corresponding to the  same values of the depth parameter ($\delta=4,\, 25, \,81$) used in the previous figure. 
First, for each family of curves we confirm the previous observations: negative vorticity increases the maximum wave age, whereas positive vorticity has the opposite effect. Interestingly, this behaviour is similar to what is observed in the absence of vorticity when the water depth is varied (see Fig.2 of ref.~\cite{Manna1}). 
Second, we observe a global enhancement of $\hat{\gamma}$ for negative vorticity, which replicates the behavior of $\beta$. However, this contrasts with the effect of water depth~\cite{Manna1}, which only displaces the maximum wave age. The enhancement of the actual growth coefficient $\hat{\gamma}$ due to vorticity is an entirely novel feature.

\subsection{Energy growth rate}

\begin{figure}[h]
  \begin{center}
    \parbox{0.6\textwidth}{
    \subfloat[$\delta=25$]{
      \includegraphics[width=0.6\textwidth]{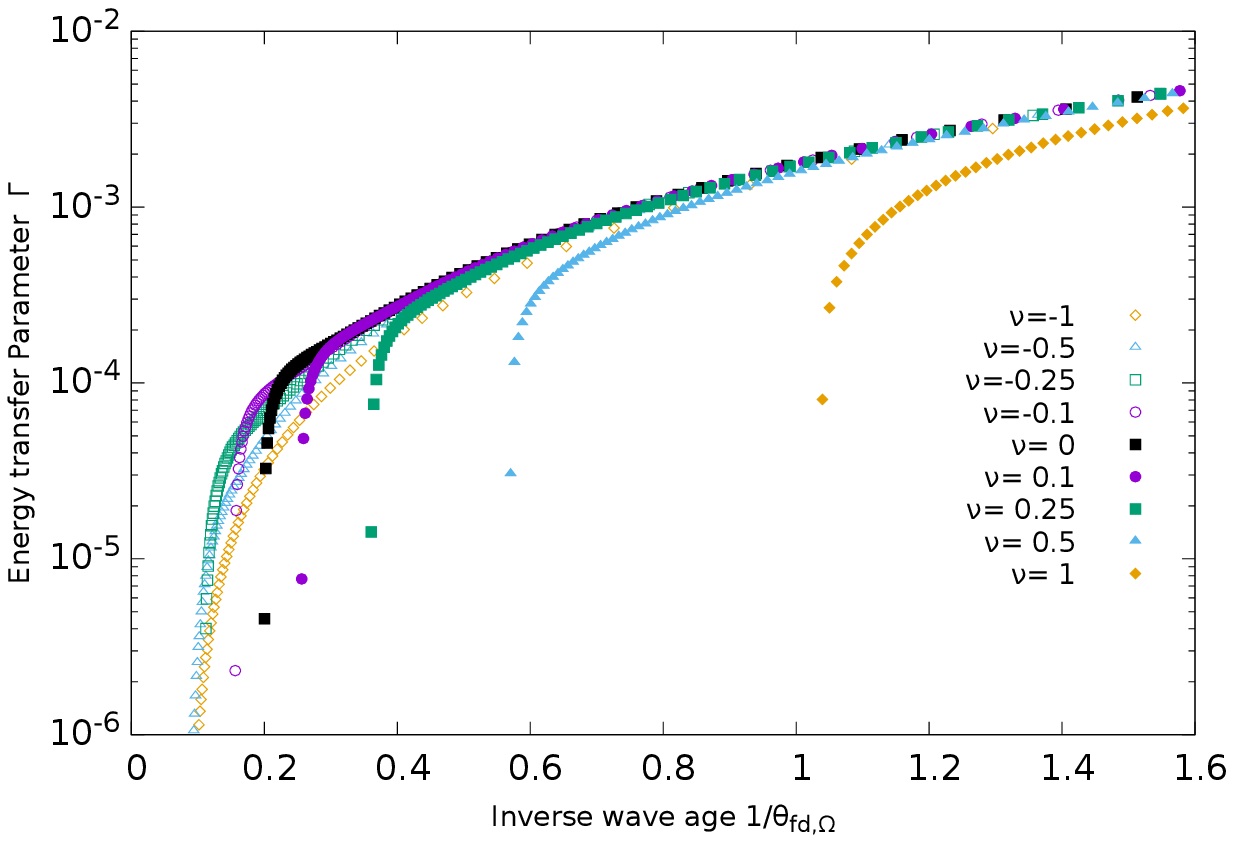}
      }
    }
    \parbox[t]{0.35\textwidth}{
    \subfloat[$\delta=9$]{
      \includegraphics[width=0.35\textwidth]{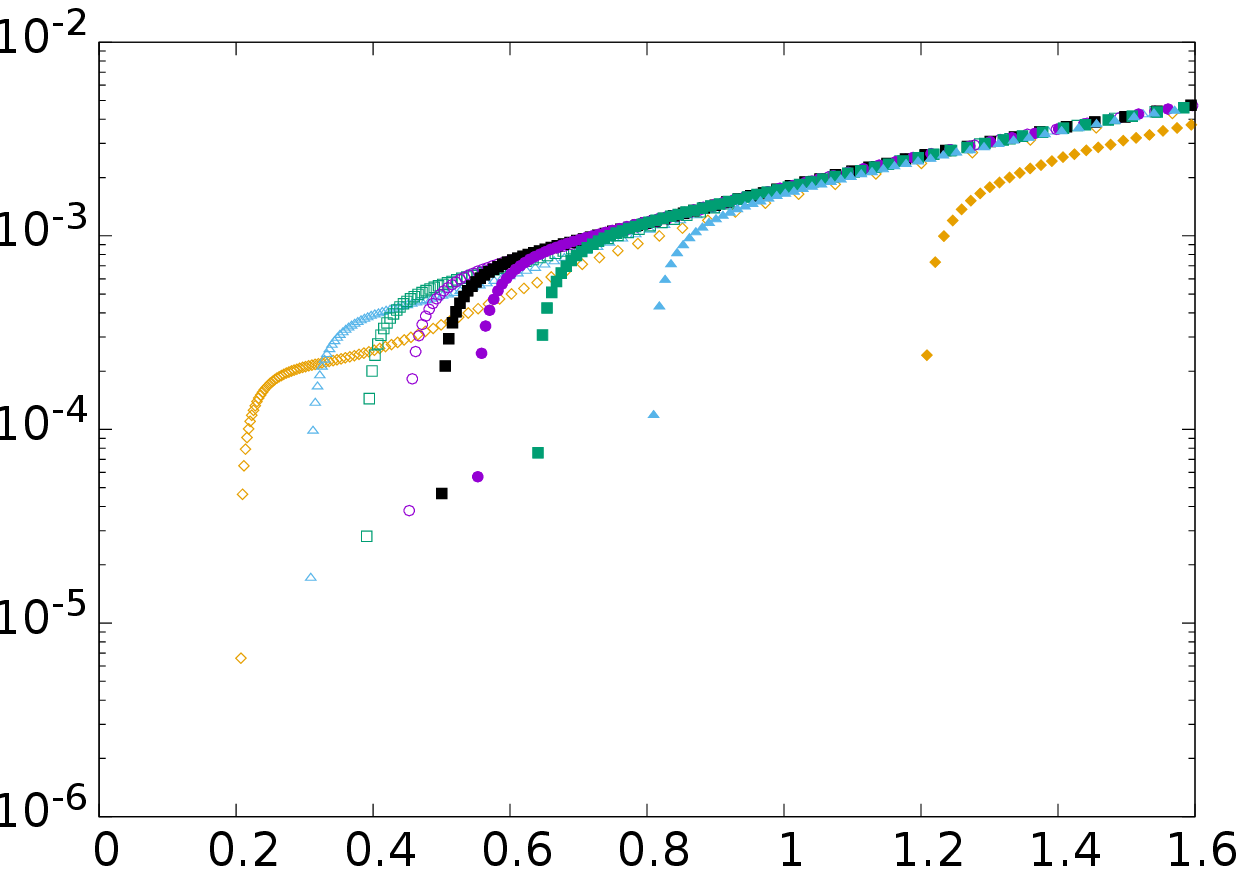}
      }
      \\
     \subfloat[$\delta=81$]{
       \includegraphics[width=0.35\textwidth]{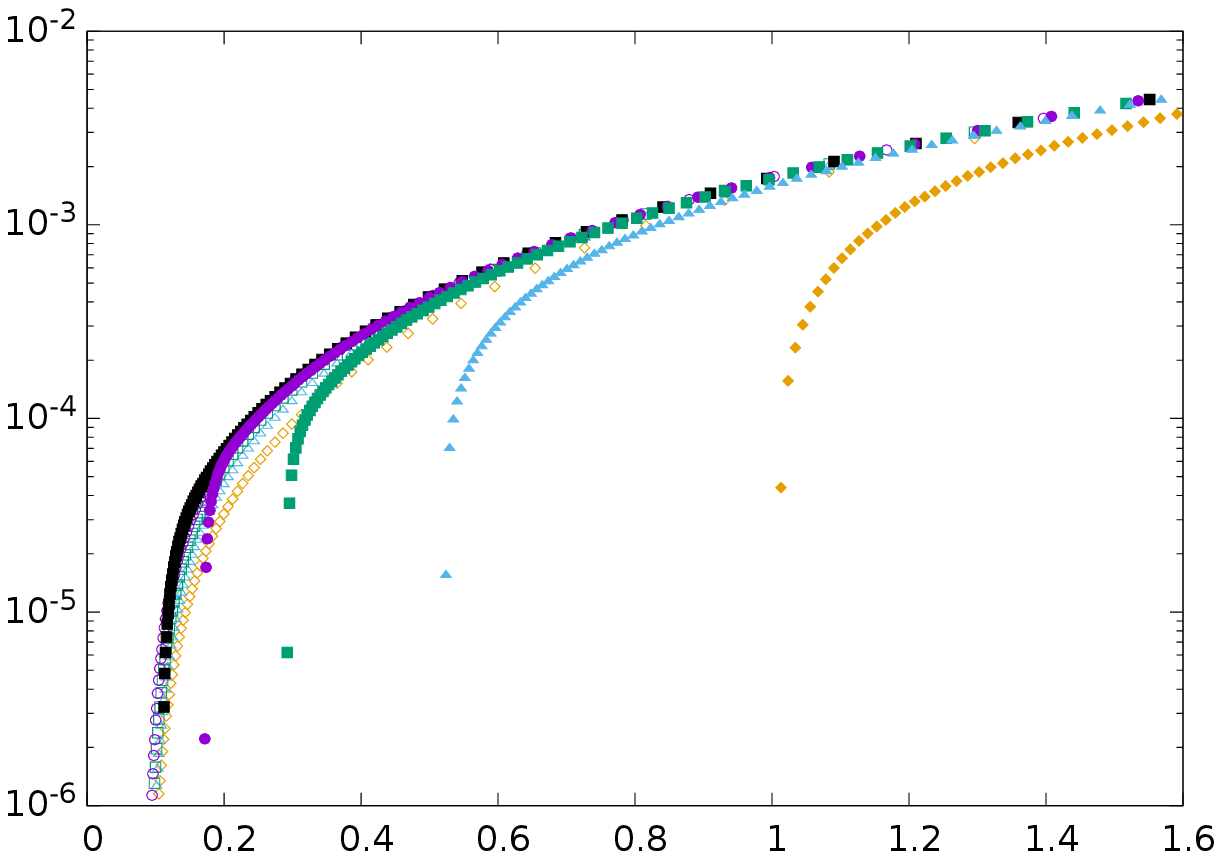}
      }
    }
    \caption{The energy transfer parameter $\Gamma$, defined by Eq.~(\ref{eq:Gamma}), as a function of the inverse wave age $1/\theta$. The choices for the depth and vorticity parameters $\delta$ and $\nu$ are as in Figs.~\ref{beta-1}, \ref{beta-2} and \ref{gamma}. In (b) and (c) the data ranges, as well as symbols, 
    are identical to those indicated in (a).
    }
    \label{Gamma-1}
  \end{center}
\end{figure}

Another commonly used way to assess the growth of waves is to represent the energy growth rate $\Gamma$ as a function of the inverse wave age $1/\theta$ , see e.g. in \cite{BejiN}. This is shown in Fig.~\ref{Gamma-1}, based on Eq.~(\ref{eq:Gamma}). Three choices for the depth parameter $\delta$ are juxtaposed, using the same values as in Figs.~\ref{beta-1}, \ref{beta-2} and \ref{gamma}. The behaviour is again reminiscent of what happens when the water depth is finite (see ref.~\cite{Manna1}), i.e. the location of the vertical drop of the energy growth rate is displaced, here under the effect of vorticity. For positive vorticity the shift is towards the right, for negative vorticity it is towards the left. This again reflects the existence of a maximum wave age, which decreases with positive vorticity but increases with negative vorticity.

On the left side of this graph, for older seas, curves do not merge and differences are well visible. Towards the right, i.e. for very young seas, all curves appear to merge asymptotically. However, small differences are still present between different graphs corresponding to different vorticities, although they are hardly visible on the logarithmic scale.  

\begin{figure}[h]
  \begin{center}
    \parbox{0.6\textwidth}{
    \subfloat[$\theta=4$]{
      \includegraphics[width=0.6\textwidth]{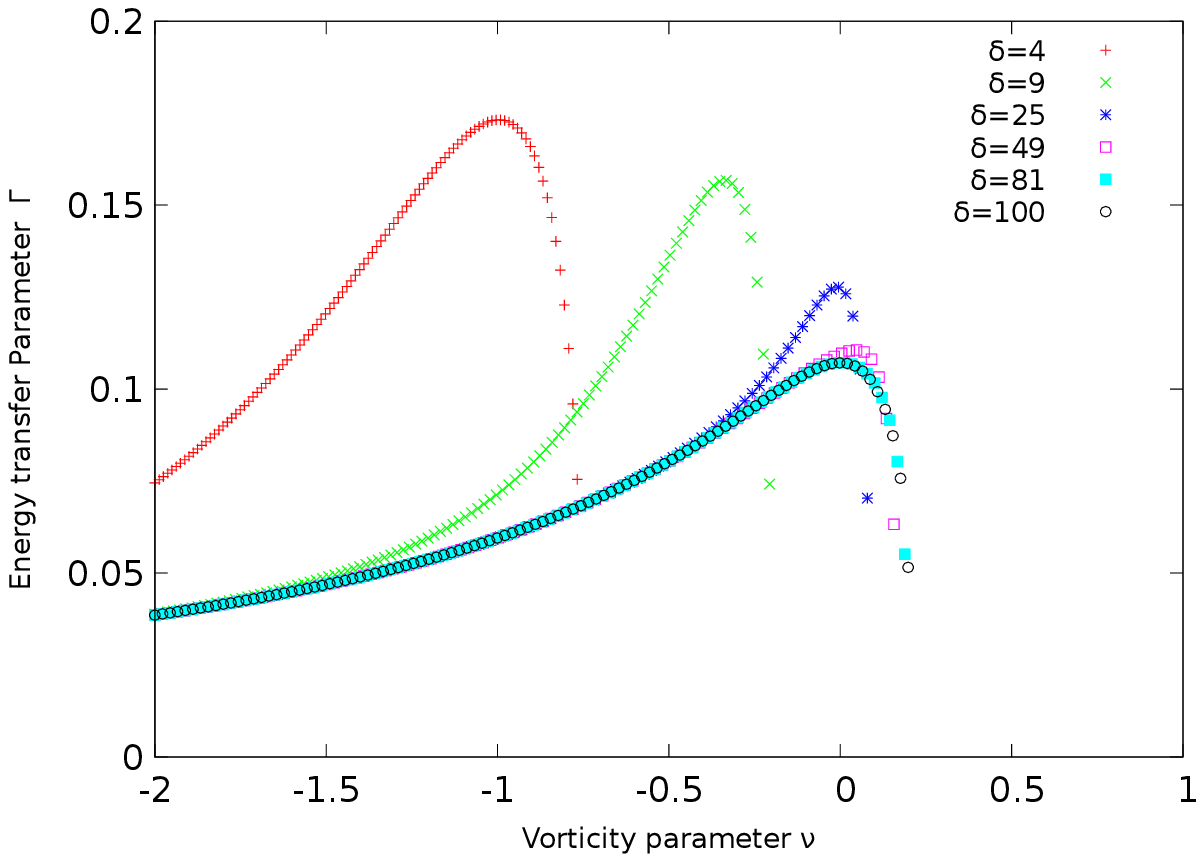}
      }
    }
    \parbox[t]{0.35\textwidth}{
    \subfloat[$\theta=1$]{
      \includegraphics[width=0.35\textwidth]{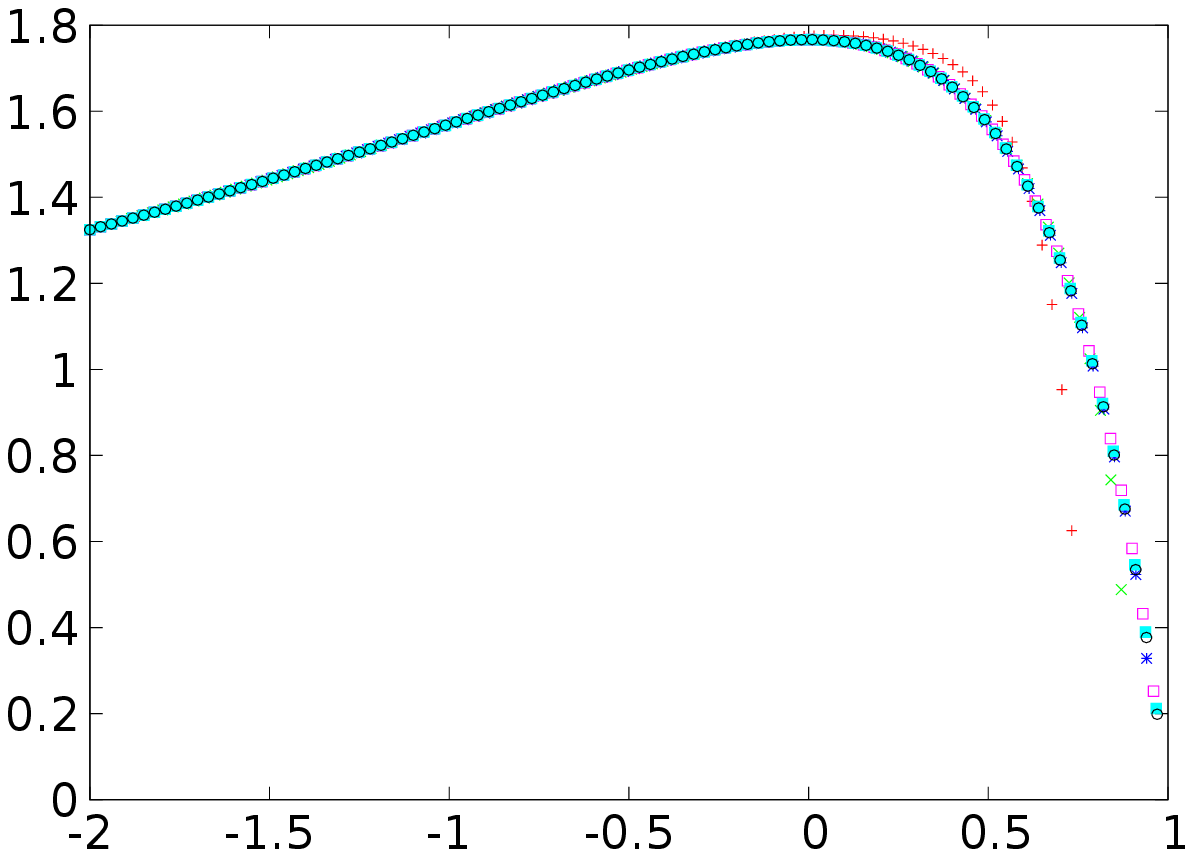}
       }
      \\
     \subfloat[$\theta=7$]{
       \includegraphics[width=0.35\textwidth]{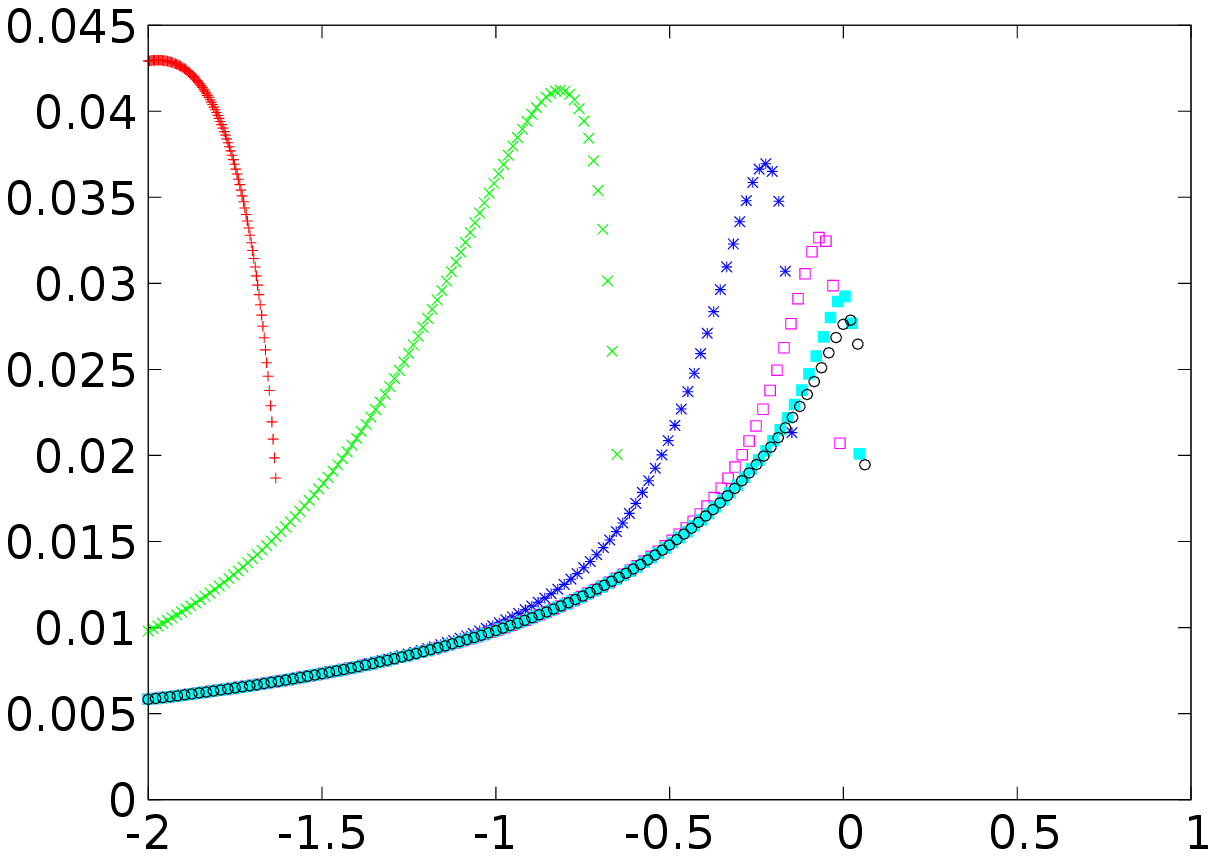}
      }
    }
    \caption{\label{Gamma-2}}
  Effect of vorticity on the energy growth parameter $\Gamma$, contrasting various choices for the wave age. Selected depth values are identical for all three panels, but note that the vertical scale is adapted for better visibility.
  For all wave ages, positive vorticity leads to a dramatic decrease in the energy growth. There always is a specific value of (negative) vorticity for which the energy growth parameter is at a maximum. 
  (a,c) For deep water, the maximum energy growth rate is achieved for zero vorticity, but this maximum progressively shifts towards stronger (negative) vorticities as the water depth is reduced.
  (b) The previous effect is not visible for very young seas: all data collapse, with a maximum at zero vorticity. Therefore vorticity of any sign reduces the wave growth in this regime.
  \end{center}
\end{figure}

An alternative view of the same quantities is therefore presented in  Fig.~\ref{Gamma-2}, where we use a linear scale. It highlights the effect of vorticity  by plotting $\Gamma (\nu)$, for three values of the wave age $\theta$ (panels a-c). For each of these, graphs for several values of the depth parameters are superposed.  

Each graph presents a maximum, i.e. there is a specific vorticity for which the energy transfer is maximum. 
In deep water it is located at zero vorticity, but occurs at negative vorticity in finite depth.
For young seas (panel (b)), the depth does not play a significant role, and the effect of vorticity is almost independent of the depth parameter $\delta$. In this case the energy transfer is maximal for zero vorticity, i.e. the presence of any shear flow will reduce the energy transfer. Therefore, as long as the sea is not well developed, vorticity of any sign will hinder wave growth. This changes as the sea develops, see panels (a) and (c). For higher wave ages, the energy growth coefficient is still maximum for a specific value of the vorticity, which is always negative in finite depth. Moreover, this value is now clearly dependent on the depth parameter $\delta$: the smaller the depth parameter, the more negative the vorticity for which the energy transfer is maximum, and the higher the energy transfer which can be achieved. Note also that, whatever the wave age, the graphs remain asymptotically universal for large depth parameters $\delta$: in deep water, the maximum energy transfer is always achieved in the absence of vorticity. Consequently, vorticity of any sign will always hinder energy transfer in deep water. 

\section{Conclusions and outlook}
\label{sec:conclusions}

We have presented a full theoretical treatment of wind-generated surface waves, on water of finite depth, when the water body is subject to a shear flow of constant vorticity. This work extends the theory of Miles to include vorticity, a topic with great practical implications since water currents are expected to be present in oceans, and may lead to particularly strong vorticity in near-coastal waters.

Generalising the approach by Beji and Nadaoka \cite{BejiN} has yielded expressions for all growth coefficients characterising the wave growth due to wind, such as the Miles growth coefficient $\beta$, the amplitude growth coefficient $\gamma$ and the energy growth coefficient $\Gamma$, known from previous studies in the absence of water currents \cite{Manna1}. 

As a first result, the notion of a generalised wave age has emerged, defined in the same spirit as in \cite{Manna1}, which accounts for the effect of both finite depth and vorticity in the water flow on the phase velocity of wind-generated waves. It shows that, in otherwise identical conditions, vorticity alters the wave age according to its sign. 
The wave age is reduced for positive vorticity, which implies that in the natural reference frame of the water surface the flow velocity is directed against the wind, as sketched in Fig. \ref{fig:sketch}.
Negative vorticity, on the other hand, leads to more developed seas.

Interestingly, there is an intrinsic limit to this wave age, i.e. an upper bound which cannot be exceeded. This is already known to arise in water of finite depth, but now the maximum wave age is modified by the presence of vorticity.  Positive vorticity reduces this maximum value, and therefore leads to 'younger' seas. Negative vorticity, however, makes it possible to attain a more developed sea.

More specifically, the state towards which the sea evolves under the effect of wind is quantified by various growth coefficients. We have determined several such coefficients, by numerically solving the Rayleigh equation, again following the strategy by Beji and Nadaoka \cite{BejiN}, which has revealed new features with respect to the case of static water.

First, we have considered the Miles growth coefficient $\beta$, related to the perturbation in the water pressure \cite{ConteMiles}. In deep water it is known to decrease exponentially to zero beyond the wave age corresponding to the developed sea \cite{BejiN,Janssen}. Recall the effect of a finite water depth: the vanishing of the Miles growth coefficient is pre-empted by the aforementioned maximum wave age, thereby reducing the wave age of the developed sea. Nevertheless, at sufficiently small wave ages the growth coefficient is unaffected by the water depth. 
The action of vorticity is now twofold. 

At a first level, increasing (positive) vorticity reduces this maximum wave age, and therefore modifies the wave age of the developed sea; negative vorticity has the opposite effect.
In this respect a change in vorticity is similar to a change in water depth. 
This implies also that a shift in maximum wave age cannot unambiguously be attributed to either finite water depth or to the presence of shear currents, as both can compensate.

On a second level, however, vorticity also affects the value of the growth coefficient for small wave ages: it is diminished for positive vorticity but enhanced for negative vorticity. This effect has no counterpart due to water depth, it is entirely novel in the present context.

The growth parameter $\hat{\gamma}$ which directly characterises the growth of the wave amplitude also mirrors this behaviour.

Finally, the energy transfer from the wind to the wave, characterised by the coefficient $\hat{\Gamma}$, shows a more complex behaviour. For young seas, we have illustrated numerically that the presence of vorticity necessarily leads to a decrease in $\hat{\Gamma}$: irrespective of the water depth, the energy transfer attains its maximum as vorticity approaches zero. For larger wave ages though, this is no longer true: the maximum in the energy transfer is achieved for a  negative value of vorticity. This value furthermore depends both on wave age and water depth.

\acknowledgements
This work has been partially funded by the MUSE program of the University of Montpellier, in the framework of the international program 'KIM Sea and Coast'.

\bibliography{References.bib}

\end{document}